\documentclass[12pt]{article}
\usepackage{xcolor}
\definecolor{darkblue}{rgb}{0,0,0.3}   

\usepackage[super,sort&compress]{natbib}

\usepackage{hyperref}
\hypersetup{colorlinks=true, citecolor=darkblue, linkcolor=darkblue, urlcolor=darkblue, hypertexnames=false}
\usepackage{caption}
\usepackage{comment}
\usepackage{float}
\usepackage{longtable}
\usepackage{booktabs}
\usepackage[english]{babel}
\usepackage{placeins}
\usepackage[letterpaper,top=2cm,bottom=2cm,left=3cm,right=3cm,marginparwidth=1.75cm]{geometry}
\usepackage{lineno}   
\usepackage{amsmath}
\usepackage{amssymb}

\usepackage{graphicx}
\usepackage{orcidlink}
\usepackage{pdfpages}
\usepackage{xcolor}
\newcommand{\rev}[1]{\textcolor{black}{#1}}

\title{\rev{Physical} disturbance expands the climatic \rev{envelope} of annual plant dominance}
\date{\today}

\makeatletter
\providecommand{\printaffiliations}{}
\makeatother

\begin{document}

{\sloppy\maketitle} 
\vspace{0.8\baselineskip}  
\vspace{0.8\baselineskip}  

\begin{abstract}
Disturbance regimes and nutrient inputs are changing worldwide, with consequences for the structure and functioning of plant communities. Life-history theory predicts that disturbance should shift communities from long-lived perennials toward short-lived annuals, and that nutrient enrichment may amplify this shift. However, these predictions have not been tested experimentally across broad environmental gradients. Here, using a global grassland experiment spanning 37 sites, we tested how \rev{severe} physical disturbance (vegetation removal and shallow soil tillage) and fertilisation \rev{alter the} annual–perennial balance, and whether disturbance \rev{expands the climatic envelope} of annual dominance. Disturbance increased annual species share \rev{from 15\% to 29\%} and relative annual cover \rev{from 9\% to 29\%}, whereas fertilisation had little influence and did not interact with disturbance. The disturbance-driven shift arose through contrasting pathways: in graminoids and legumes, \rev{perennial cover declined}, while in forbs, \rev{annual cover expanded}. In the absence of disturbance, annual dominance was restricted to systems with extremely hot and dry summers, \rev{yet disturbance consistently increased annual prevalence across climatic gradients and doubled the extent of climate space in which annuals dominated}. \rev{During repeated disturbance,} annuals may help maintain vegetation cover, but their expansion also signals loss of perennial cover and \rev{associated ecosystem functions}.

\end{abstract}

\newpage

\section{Introduction}

\noindent
Land-use intensification is transforming grasslands worldwide through increased disturbances and nutrient inputs, with consequences for vegetation structure and ecosystem functioning \citep{foley_global_2005,gossner_landuse_2016}. These changes may reorganize grassland communities by shifting the balance between annual and perennial life-history strategies. Annuals represent the short-lived extreme of plant life span, completing their life cycle within a year, while perennials persist for longer and often reproduce repeatedly \citep{friedman_all_2015}. Hence, annuals often allocate more to rapid aboveground growth and seed production, while perennials tend to invest more in storage organs and belowground structures \citep{vico_tradeoffs_2016}. Because perennials generally contribute more to long-term carbon storage, soil stabilisation, and hydrological regulation \citep{glover_increased_2010,vico_tradeoffs_2016}, shifts in annual--perennial balance can have important consequences for ecosystem functioning.

Annual species constitute $\sim$15\% of the world’s herbaceous flora, with substantially higher representation in hot, dry climates \citep{poppenwimer_revising_2023}. They also dominate many human-altered and agricultural systems \citep{pimentel_annual_2012,clark_contingent_2019,vile_structural_2006}. Classical life-history theory provides a demographic basis for this pattern: annuals should dominate when adult survival is low relative to juvenile survival \citep{cole_population_1954,pianka_rand_1970,charnov_lifehistory_1973,grime_plant_1979,pacala_models_1998}. \rev{Physical} disturbance can create these conditions by reducing survival of established individuals, particularly long-lived perennials, while opening recruitment windows for species that regenerate rapidly from seeds. Annuals are often over-represented in persistent soil seed banks \citep{demalach_contrasting_2023}, may disperse seeds effectively \citep{beckman_high_2018}, and can benefit from litter removal and newly opened patches after disturbance \citep{grubb_maintenance_1977,crawley_population_1987,paniw_interacting_2017,pacala_models_1998}. \rev{Yet, disturbance need not always favor annuals, as long-distance clonal spread can provide some perennials with an alternative strategy for rapidly occupying disturbed patches \citep{fahrig_advantage_1994}.}

\begin{figure}[H]
    \centering
    \includegraphics[width=1\linewidth]{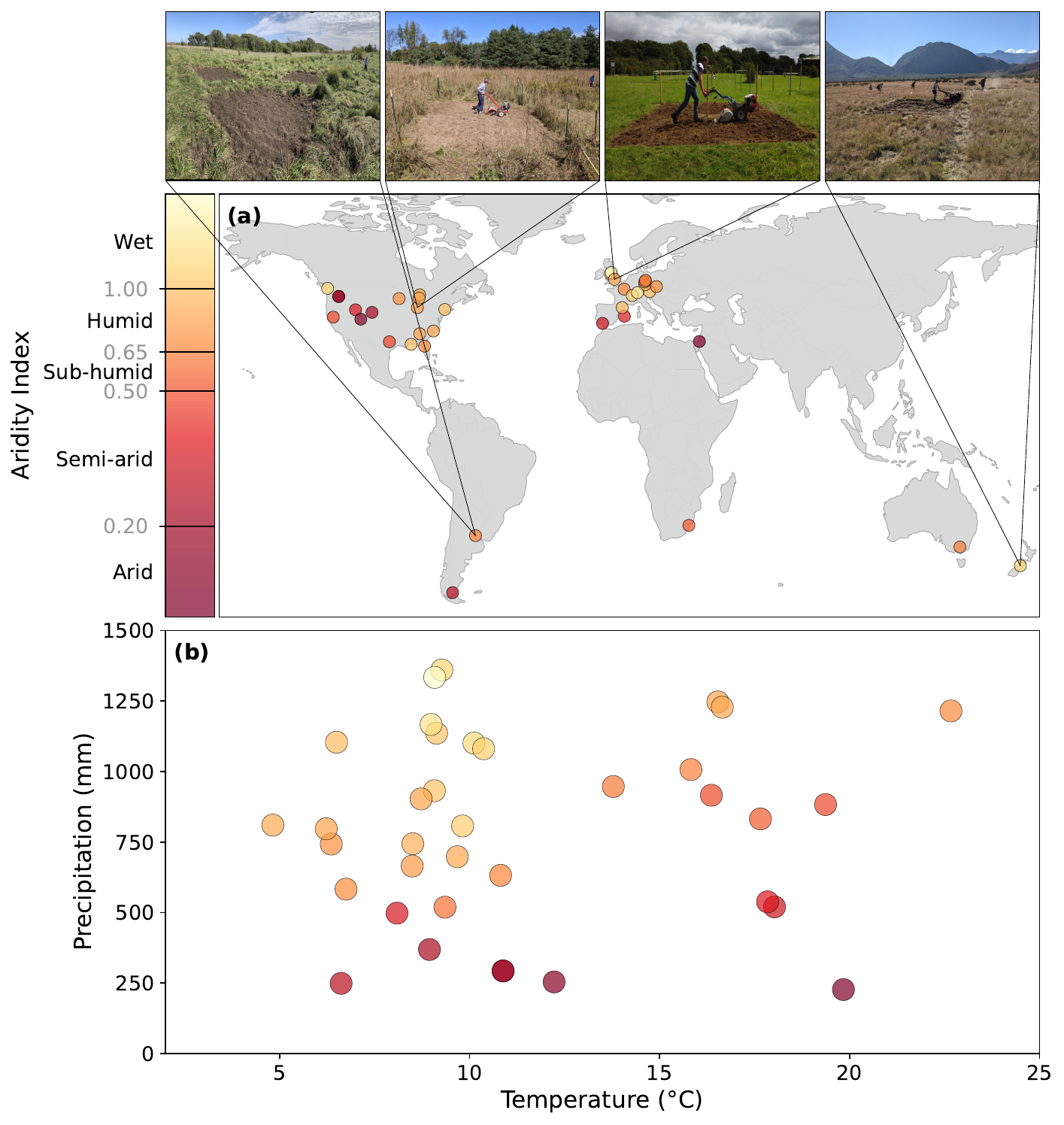}
\caption{
\textbf{Geographic and climatic distribution of the study sites.}
The DRAGNet experiment applied coordinated disturbance (vegetation removal and tilling) and fertilisation treatments across grassland sites spanning a broad climatic gradient.
Photographs at the top show examples of DRAGNet grassland sites from different continents.
\textbf{(A)} Geographical distribution of the 37 sites included in this study, colored by site aridity, with redder colors indicating more arid conditions.
\textbf{(B)} Study sites plotted across mean annual temperature and annual precipitation.
Photographs are shown from left to right: Argentina (San Claudio; photo credit: Pedro Tognetti), United States (Kellogg Biological Station; photo credit: Lars Brudvig), United Kingdom (Wytham Woods; photo credit: Roberto Salguero-Gomez), and New Zealand (Cass; photo credit: Andrew Barnes).
Additional site information is provided in Table~\ref{table:sites}.
}
    \label{fig:map}
\end{figure}

Fertilisation could also favor annuals, because they often exhibit traits associated with rapid resource acquisition, such as high specific leaf area and fast growth, potentially allowing them to complete their life cycle before subsequent disturbance or climatic stress reduces survival \citep{westoby_plant_2002, vico_tradeoffs_2016}. \rev{Conversely, fertilisation could favor perennials through} their greater capacity to capture and retain resources through time, together with their vegetative growth capacity, which enables them to maintain dominance under high nutrient availability \citep{tilman_plant_1988, gough_incorporating_2012}.

Local-scale experiments have \rev{examined the annual-perennial balance} under contrasting environmental conditions, particularly during post-disturbance succession and community reassembly \citep{seabloom_invasion_2003, shipley_plant_2006, perez-camacho_plant_2012, clark_contingent_2019}. At the other extreme, large-scale observational studies have linked the global distribution of annual strategies to climatic gradients using species occurrence data \citep{poppenwimer_revising_2023, taylor_contribution_2023,boyko_evolutionary_2023}. While these approaches have clarified biotic mechanisms at small scales and climatic correlates at broad scales, they have rarely been integrated. In particular, we still lack experimental tests of whether disturbance and fertilisation systematically alter annual–perennial balance across climate gradients, and whether such effects can expand the \rev{climatic envelope of annual dominance}.

Here, we test how severe physical disturbance (vegetation removal and \rev{shallow} soil tillage\rev{, applied annually for three consecutive years}), fertilisation, and their interaction shape the prevalence of annual plants, quantified as relative annual cover and annual species share (the proportion of species that are annual). We further ask \rev{whether annual prevalence and disturbance responses vary across climatic gradients. Because grassland functional groups differ in rooting, nutrient-acquisition, and vegetative-regeneration strategies and often respond differently to disturbance and fertilisation \citep{demalach_contrasting_2017, nelson_forb_2025,tognetti_negative_2021}, we also ask whether shifts in annual–perennial balance are general across graminoids, forbs, and legumes or driven by particular groups.} To address these questions, we use the coordinated Disturbance and Recovery Across Global Grasslands Network (DRAGNet \citep{borer_maximizing_2025}), which applies common experimental treatments and vegetation sampling protocols across 37 grassland sites (Fig.~\ref{fig:map}).

Based on life-history theory, we expected disturbance to favor annuals over perennials, \rev{with no consistent directional expectation for fertilisation}. \rev{We expected a positive disturbance~$\times$~fertilisation interaction, implying that} nutrient addition \rev{would} amplify disturbance-driven increases in annuals. Finally, we expected baseline annual dominance to be \rev{restricted mainly to} hot, dry, and highly seasonal environments, and disturbance to \rev{expand the climatic envelope in which annuals dominate}.

\noindent

\section{Results}

\begin{figure}[H]
\centering
\includegraphics[width=0.8\columnwidth]{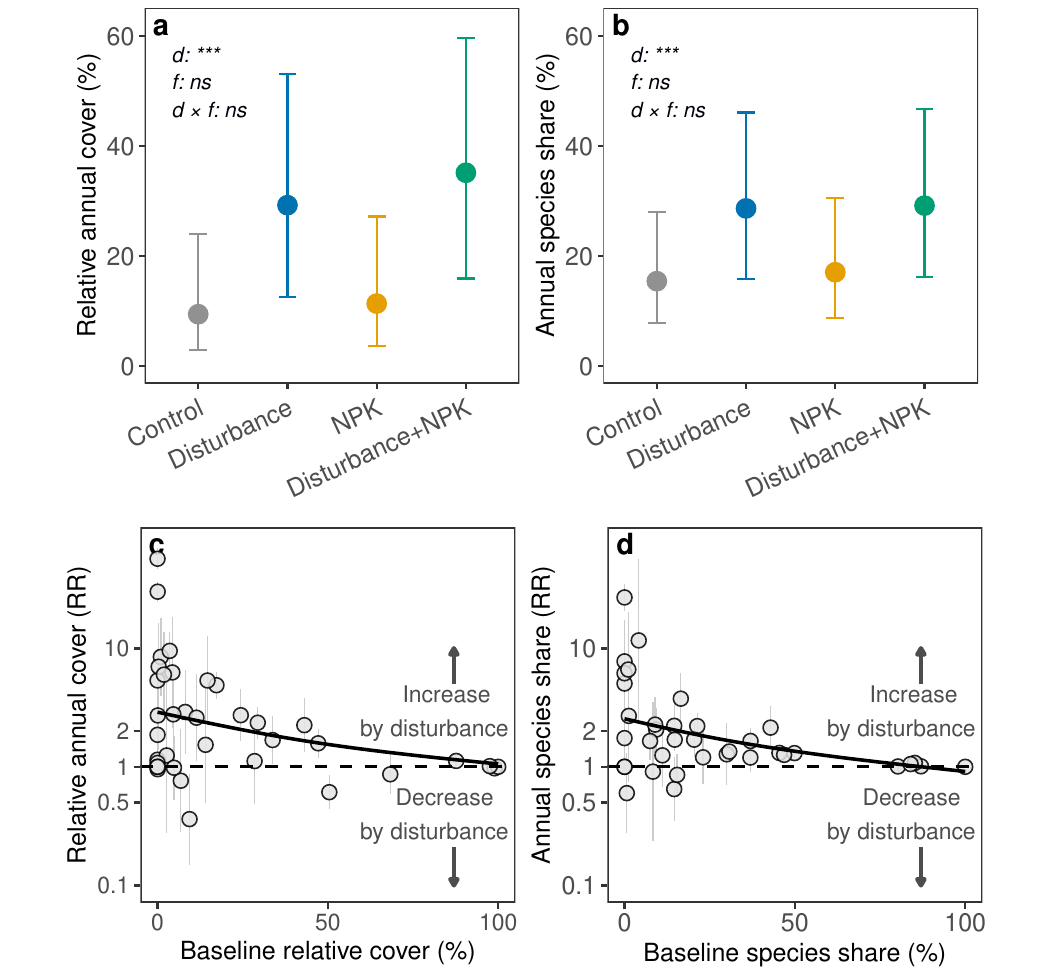}
\caption{
\textbf{\rev{Physical} disturbance consistently increases the prevalence of annuals, whereas fertilisation has no effect.}
\textbf{Panels A--B: network-level treatment effects.}
\textbf{(a)} Relative annual cover, defined as the proportion of total vegetated cover contributed by annual species.
\textbf{(b)} Annual species share, defined as the proportion of observed plant species that were annual.
Points show estimated marginal means with 95\% confidence intervals from mixed-effects models across four treatment groups.
Relative annual cover was modelled using \rev{an ordered} beta regression, and annual species share using a binomial regression
(Tables~\ref{tab:zibetaregression},~\ref{tab:logisregression}).
Annotations show planned factorial tests of the two experimental drivers and their interaction:
d = disturbance, f = \rev{NPK} fertilisation, and d~$\times$~f = disturbance~$\times$~fertilisation interaction
(***~$p<0.001$; ns~$p>0.05$). $n = 1{,}444$.
\textbf{Panels C--D: site-level disturbance effects.}
\textbf{(c)} Relative annual cover.
\textbf{(d)} Annual species share.
Each point represents one site. Baseline values on the x-axis are site-level means in control plots.
The y-axis shows the disturbance response ratio (RR, Disturbance/Control), calculated on a log scale and back-transformed for display. Solid lines show Gaussian GAM smoothers fitted to log response ratios using penalized \rev{cubic regression} splines with the basis dimension ($k = 5$) and generalized cross-validation. Approximate F-tests indicated non-zero smooth terms for relative annual cover ($p = \rev{0.022}$) and annual species share ($p = 0.008$). $n = 37$ sites.}
\label{fig:treatment_fig}
\end{figure}

\rev{Physical} disturbance markedly increased the prevalence of annuals, raising relative annual cover \rev{from 9\% to 29\%} and annual species share from 15\% to 29\% (Fig.~\ref{fig:treatment_fig}a,b; Tables~\ref{tab:zibetaregression}--\ref{tab:logisregression}). \rev{Both annual and perennial components contributed to these shifts: disturbance decreased perennial cover and richness while increasing annual cover and richness} (Tables~\ref{tab:neg_binomial_annual}--\ref{tab:zi_lognormal_perenn}).

Disturbance increased annual prevalence in nearly all sites, although the magnitude of the response was highly variable (Fig.~\ref{fig:treatment_fig}c,d). Part of this variation was associated with baseline annual values in control plots, with the largest proportional gains occurring where annuals were initially rare. \rev{Fertilisation, in contrast, tended to increase plant cover and reduce species richness, but these changes did not consistently favour annuals over perennials (Tables~\ref{tab:neg_binomial_annual}, \ref{tab:neg_binomial_perenn}, \ref{tab:zi_lognormal_annual}, and~\ref{tab:zi_lognormal_perenn}). Consequently, }fertilisation had little influence on annual prevalence, either alone or in interaction with disturbance, and showed weak and inconsistent site-level effects (Fig.~\ref{fig:treatment_fig}a,b; Appendix Fig.~\ref{fig:npk_multi}).

\begin{figure}[H]
\centering
\includegraphics[width=1\columnwidth]{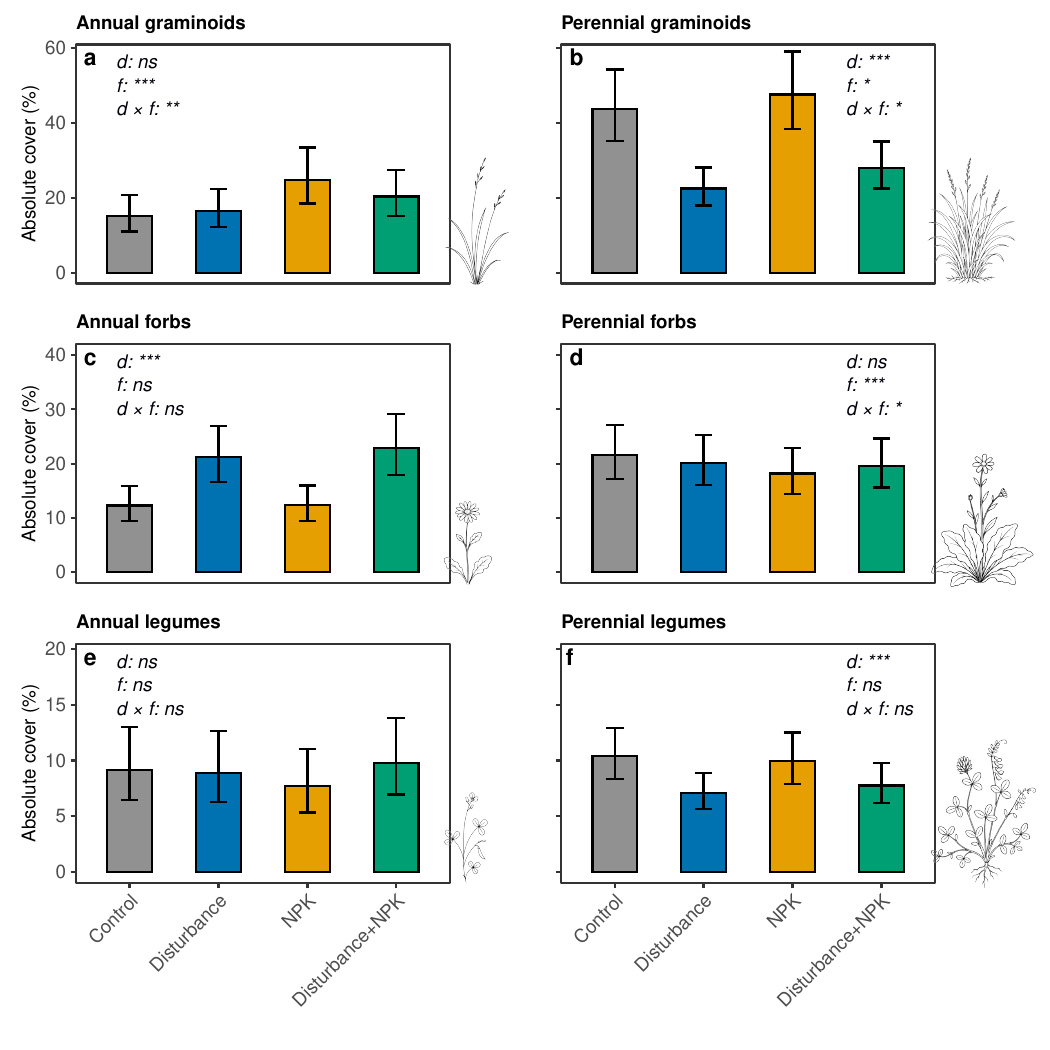}
\caption{
\textbf{\rev{Physical} disturbance shifts annual--perennial balance through different functional-group pathways.}
Bars show model-based marginal means of absolute cover with 95\% confidence intervals, estimated from the continuous component of zero-inflated lognormal mixed-effects models. \rev{Thus, for each functional group, estimates represent predicted cover conditional on its presence, rather than mean cover across all plots.}
Panels show \textbf{(a)} annual graminoids, \textbf{(b)} perennial graminoids, \textbf{(c)} annual forbs, \textbf{(d)} perennial forbs, \textbf{(e)} annual legumes, and \textbf{(f)} perennial legumes. Note the different y-axis scales among rows.
Annotations show planned factorial tests of the two experimental drivers and their interaction for the same model component: d = disturbance, f = fertilisation, and d~$\times$~f = disturbance~$\times$~fertilisation interaction \rev{(***~$p < 0.001$; **~$p < 0.01$; *~$p < 0.05$; ns~$p > 0.05$).}
\rev{Full models (including zero inflation components)} are reported in Tables~\ref{tab:table1_fig2}--\ref{tab:table6_fig2}. \rev{$n = 1{,}444$}.
}

\label{fig:functional_groups}
\end{figure}

\rev{Within functional groups, disturbance shifted annual--perennial balance through different pathways. Figure~\ref{fig:functional_groups} shows changes in cover where each group was present, while Tables~\ref{tab:table1_fig2}--\ref{tab:table6_fig2} report both conditional cover and the probability of zero cover. In graminoids and legumes, perennial cover declined while annual cover showed no clear response. In forbs, perennial cover showed no clear response to disturbance, while annual forb cover almost doubled. Changes in occurrence reinforced these patterns for the two strongest responses: disturbance increased the occurrence of annual forbs and decreased that of perennial graminoids. Under fertilisation, graminoid cover increased, perennial forb cover declined, and legumes showed no consistent response.}

\begin{figure}[H]
\centering
\includegraphics[width=1.\columnwidth]{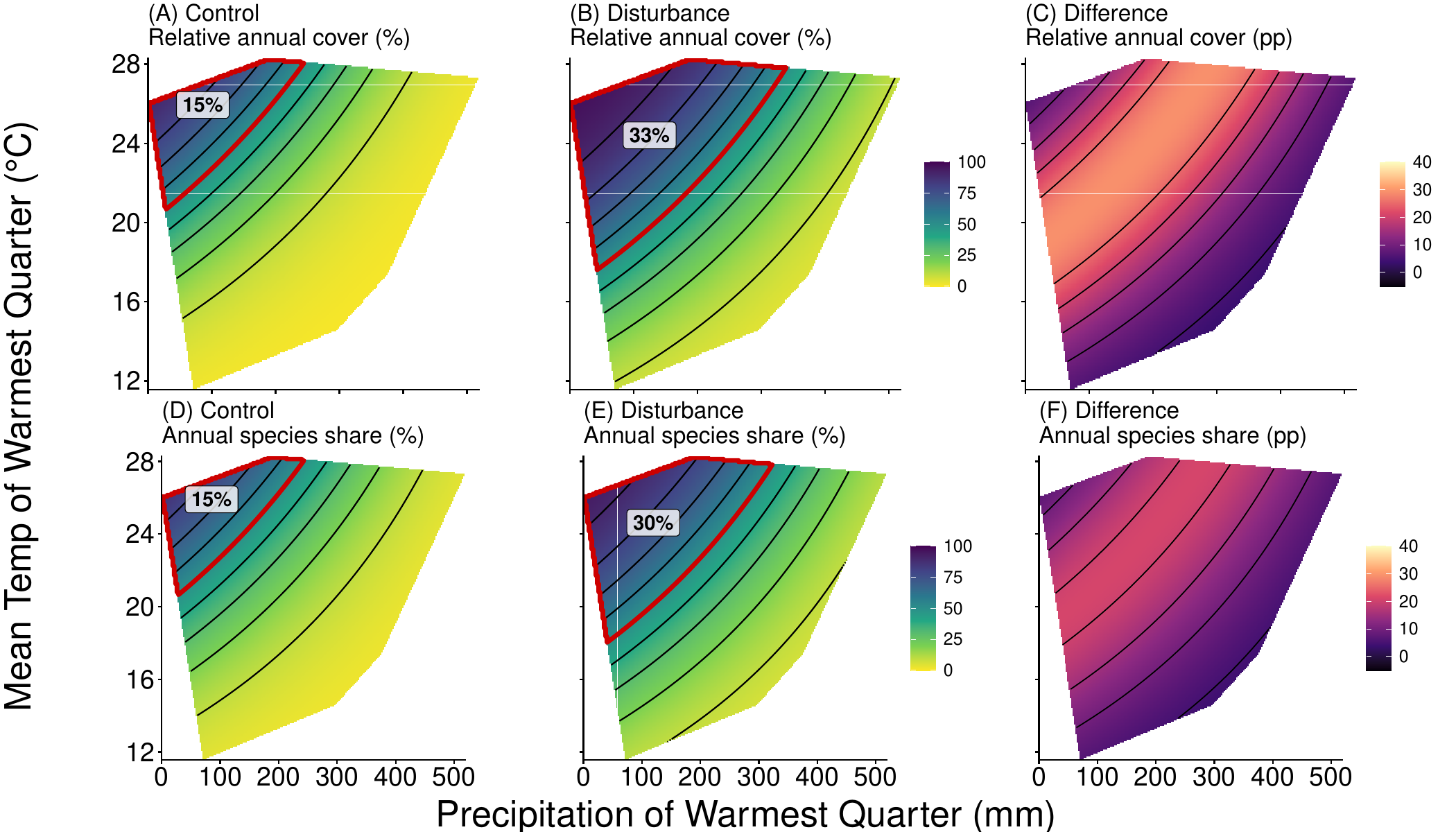}
\caption{%
\textbf{\rev{Physical} disturbance expands the climatic \rev{envelope} of annual plant dominance.}
Panels show modelled annual prevalence across mean temperature ($^\circ$C) and precipitation (mm) of the warmest quarter, \rev{predicted from ordered beta mixed-effects models for relative annual cover and annual species share}. Panels A--C show relative annual cover (\% of cover), and panels D--F show annual species share (\% species). In each case, the first two columns show predicted values under Control and Disturbance, with red solid lines marking the \rev{boundary of the climatic envelope} of annual dominance (predicted values \textgreater{}50\%). Labels inside the red boundary indicate the percentage of climate space occupied by the annual-dominance zone. Panels C and F show \rev{the predicted absolute difference between Disturbance and Control}, expressed in percentage points (pp). White areas indicate unsampled climates. \rev{Observed site locations within the climatic space, together with their annual-prevalence values, are shown in Figs.~\ref{fig:sim_unfert_cover} and~\ref{fig:sim_unfert_spec}. $n = 74$ site~$\times$~treatment observations.}
}

\label{fig:climate_fig4}
\end{figure}

To link the disturbance responses to climate, we modelled annual prevalence across the temperature--precipitation space represented by the network. Mean temperature and precipitation in the warmest quarter were the main climatic \rev{predictors} of variation in baseline annual prevalence (Tables~\rev{\ref{tab:climate_cover_combined}--\ref{tab:climate_long_table}}). Across the climatic space defined by these variables, annual dominance (defined as \textgreater{}50\% \rev{relative cover or species share}) was restricted to a narrow zone characterized by the highest temperatures and lowest warmest-quarter precipitation in the absence of disturbance (Fig.~\ref{fig:climate_fig4}). However, disturbance markedly expanded this zone: the \rev{estimated} proportion of climatic space dominated by annuals increased \rev{from 15\% to 33\% when measured by relative cover, and from 15\% to 30\% when measured by species share} (Fig.~\ref{fig:climate_fig4}). 

\rev{Despite the strong climatic association with baseline annual prevalence, disturbance increased annual prevalence across the entire climatic space, with no statistically supported disturbance~$\times$~climate interaction on the model's logit scale (Tables~\ref{tab:climate_interaction_lrt},~\ref{tab:climate_interaction_wald_cover},~\ref{tab:climate_interaction_wald_species}). Nevertheless, absolute disturbance effects varied substantially across climate space (Fig.~\ref{fig:climate_fig4}c,f), with the largest changes generally occurring where baseline annual prevalence was intermediate.}

\noindent

\FloatBarrier          

\FloatBarrier
\section{Discussion}
Ecological theory predicts that disturbance shifts annual--perennial balance toward short-lived annual strategies \citep{cole_population_1954,pianka_rand_1970,charnov_lifehistory_1973,grime_plant_1979,pacala_models_1998,demalach_alternative_2021,salguero-gomez_fastslow_2016}. Consistent with this prediction, \rev{severe physical} disturbance increased both annual species share and relative annual cover, while nutrient enrichment had minor and inconsistent effects.

\rev{Across climatic gradients, disturbance had a consistent positive effect on the model's logit scale, but the resulting absolute increases in annual prevalence varied markedly. These increases expanded the climatic envelope of annual dominance beyond the extremely hot and dry conditions to which it was otherwise restricted}. Together, these results identify \rev{physical} disturbance as a key driver of annual dominance across grasslands, capable of transforming both community structure and the climatic geography of plant life-history strategies.

\subsection{\rev{Physical} disturbance favors annuals, whereas nutrient enrichment does not}

The disturbance applied in this study removed nearly all aboveground biomass and disrupted shallow belowground structures, creating the severe reduction in adult persistence under which annual strategies are expected to gain an advantage. Under such conditions, reassembly likely depends primarily on colonization from outside the plot, supplemented by recruitment from seeds that survived the disturbance treatment. Annual plants are often more effective colonizers of newly opened patches \citep{grubb_maintenance_1977,crawley_population_1987} and are disproportionately represented in persistent seed banks \citep{demalach_contrasting_2023}. These traits allow annuals to exploit recruitment windows immediately following disturbance \citep{pacala_models_1998}, while many perennials rely on vegetative persistence \citep{klimesova_belowground_2025} that was \rev{likely disrupted} by the disturbance treatment. Together, these processes favor annuals while disadvantaging perennials after disturbance.

In contrast, nutrient enrichment did not systematically favor annuals and \rev{there was no disturbance~$\times$~fertilisation interaction,  providing no support for an amplifying effect of fertilisation by accelerating post-disturbance growth and competitive turnover, as predicted for fast, opportunistic strategies \citep{pianka_rand_1970}. The absence of interaction may suggest that, in this experiment, disturbance was the stronger driver of changes in annual–perennial balance over the period examined while fertilisation effects may emerge over longer timescales than those captured here \citep{tian_nutrient_2025}}.

Nonetheless, our results suggest that these drivers act through different mechanisms. Disturbance resets communities by removing established vegetation and \rev{could} shift assembly toward recruitment by dispersal, whereas fertilisation influences resource acquisition during subsequent growth, where advantages of fast-growing annuals may be offset by the size, longevity, and clonal growth of perennials \citep{tilman_plant_1988,gough_incorporating_2012}.

\subsection{Consistency and heterogeneity across functional groups, sites, and climatic gradients}

Disturbance shifted annual--perennial balance toward annuals across functional groups, but the pathways underlying this shift varied. Forbs, the most diverse group in grasslands \citep{brathen_paradox_2021}, are widely known to increase following disturbance, \rev{which may create recruitment opportunities for species otherwise inhibited by dense graminoid canopies \citep{nelson_forb_2025}. Our results indicate that this response was driven mainly by annual forbs, while perennial forb cover remained relatively stable. This network-level response integrates variation across many species and sites, so the relative stability of perennial forbs may partly reflect contrasting responses that average out. Clonal spread represents one potential mechanism by which some perennial species can persist or recolonize under frequent disturbance \citep{fahrig_advantage_1994}, although we lack standardized data on clonality across the network and therefore cannot evaluate its contribution to this pattern. In contrast with forbs, graminoids and legumes showed declines in perennial cover without clear changes in annual cover. In graminoids, which often dominate in the absence of disturbance \citep{brathen_paradox_2021}, this decline may reflect the reliance of perennials on belowground organs for persistence and regrowth \citep{stromberg_history_2022}, which are particularly vulnerable to soil disruption. The parallel decline in perennial legumes has potential implications for nitrogen cycling and forage quality \citep{tognetti_negative_2021}.}

\rev{We found substantial variation among sites on both proportional and absolute scales, but disturbance effects were broadly consistent across climatic gradients on the model's logit scale. This suggests that much of the systematic variation in effect size reflects differences in baseline annual prevalence across climates and the bounded nature of the response, although variation among sites with similar baselines points to additional local factors not captured here. Importantly, variation in absolute effect size across climate space may be ecologically important, because larger shifts in annual cover could translate into larger changes in ecosystem functioning.}

\subsection{\rev{Physical} disturbance expands the climatic \rev{envelope} of annual dominance}

In the absence of disturbance, annual dominance was restricted to hot climates with very low precipitation during the warmest quarter. Previous biogeographic and evolutionary analyses \rev{documented this climatic pattern} for annual species share \citep{poppenwimer_revising_2023,taylor_contribution_2023,boyko_evolutionary_2023}; here, we show that the same \rev{pattern also characterizes} relative annual cover, the metric more directly linked to ecosystem functioning. If only species share had varied with climate, the pattern could reflect compositional turnover without dominance shifts; if only relative cover had varied, it could reflect dominance by only a few annual species. Their joint response therefore supports the interpretation that the climatic \rev{association} extends to the annual--perennial life-cycle axis itself.

Repeated physical disturbance markedly expanded the climate space in which annual dominance occurs, allowing annuals to dominate across a much broader range of temperature and precipitation conditions. \rev{Thus, the observed climatic envelope of annual dominance depends strongly on disturbance regime, not on climate alone.}

\subsection{Conclusions}

Annual plants are often viewed as transient or subordinate components of grassland ecosystems\citep{clark_contingent_2019,pacala_models_1998,tilman_plant_1988}. \rev{Our results instead show that repeated physical disturbance can shift annual--perennial balance consistently across grasslands worldwide and expand the climatic envelope in which annual strategies dominate. The contrasting responses of annual and perennial cover further suggest an important role for annuals in the resistance of vegetation cover to repeated disturbance: annual cover was maintained or increased within each functional group, whereas perennial cover declined in graminoids and legumes. Annuals may therefore partially compensate for disturbance-induced losses of perennial vegetation and help sustain plant cover during repeated disturbance.}

\rev{At the same time, resistance of vegetation cover does not imply functional equivalence between annuals and perennials.} Perennial dominance and extensive belowground biomass are associated with many ecosystem services provided by grasslands, including long-term carbon storage, erosion control, and regulation of water runoff\citep{vico_tradeoffs_2016,glover_increased_2010}. \rev{Thus, the expansion of annuals may help buffer losses of vegetation cover during repeated disturbance while simultaneously signalling erosion of perennial dominance and the longer-term ecosystem functions associated with it. Our findings advance understanding of the broad range of conditions under which physical disturbance promotes annual dominance, helping to predict and manage grassland responses to global change.}

\section*{Methods}

\noindent
\textbf{Experimental design and vegetation sampling.}
These data were generated as part of the Disturbance and Recovery Across Global Grasslands Network (DRAGNet). DRAGNet is a recently established global experiment explicitly designed to test the effects of physical disturbance, alone and in combination with nutrient enrichment, across grassland ecosystems worldwide \citep{borer_maximizing_2025}.  DRAGNet sites span a broad climatic gradient, from mean annual temperatures of approximately $-5$ to $25^{\circ}\mathrm{C}$ and annual precipitation from roughly $200$ to $1700\,\mathrm{mm}$, and include both natural and semi-natural grasslands \citep{borer_maximizing_2025}.

At each site, DRAGNet follows a randomized block design with 3--5 blocks. Each block contains one $5 \times 5 \mathrm{m}$ plot assigned to each of four treatments used here: an unmanipulated Control, Disturbance, Nutrient Addition (NPK), and their combination (Disturbance + NPK). A fifth Nutrient Cessation treatment is part of the DRAGNet design but not analysed in this study. \rev{Nutrient addition consists of annual applications of $10\,\mathrm{g\,m^{-2}}$ each of elemental nitrogen, phosphorus, and potassium, with micronutrients added once during the first treatment year.} The Disturbance treatment is applied once per year for three consecutive years and is designed to create conditions for community reassembly by removing established vegetation and disrupting shallow belowground structures. Specifically, all standing biomass is removed, large visible stolons and rhizomes are extracted, and the soil is tilled to a target depth of 15 cm, or to rooting depth or the rock layer in shallow soils. This treatment approximates severe physical land-use disturbances, such as tillage or mechanical vegetation removal, that remove established plants and open space for recolonization.

Within each $5 \times 5\,\mathrm{m}$ plot, vegetation is sampled once per year during the growing season in a permanently marked $1\,\mathrm{m}^2$ quadrat located at least $1\,\mathrm{m}$ inside the plot boundaries. For each vascular plant species, observers visually estimate percentage cover, allowing total cover to exceed $100\%$ where canopies overlap. Vegetation is surveyed after plants have had time to re-establish following the annual disturbance; in disturbed plots, annual and perennial plants therefore establish either by dispersal or from below-ground organs and seeds that \rev{survived the disturbance, including seeds below the tilled layer}. We analysed \rev{up to} the first three years after treatment initiation at each site, corresponding to the period during which the annual disturbance treatment was actively applied. Treatment effects therefore represent average differences between treated and control plots during this active treatment period, rather than responses to a single disturbance event. \rev{Because sites contributed only two to three active-treatment years, we focused on the average effect across this period rather than temporal trends, which over such short series are difficult to separate from year-to-year variability and may differ among sites.} After applying a small set of a priori filters (Appendix~\ref{appendixs1}), the dataset comprised 37 sites and 1,444 plot--year observations. \rev{Response variables were calculated separately for each plot in each year, and all retained plot--year observations from the two to three active-treatment years were included in the network-wide models.}

\medskip
\noindent
\textbf{Response variables.}
We quantified the prevalence of annuals at the plot level using two complementary metrics: (i) the proportion of annual species out of total species richness (species share), and (ii) the proportion of total vegetated cover contributed by annuals (relative annual cover). Species share captures the \rev{representation of annuals in species composition}, while relative annual cover \rev{captures their contribution to total vegetated cover}. In addition, we analysed richness and absolute cover (\% cover) of annuals and perennials, and analogous responses within functional groups (graminoids, forbs, legumes), to understand which components of the community drive the aggregate patterns (Appendix~S1). \rev{The analyses therefore use different data representations for distinct questions: plot--year responses for network-wide factorial treatment effects, site-level response ratios for among-site variation in disturbance effects, and site~$\times$~treatment means for the climate analyses.}

\medskip
\noindent
\textbf{Network-wide treatment effects.}
To estimate the average effects of disturbance and fertilisation across all sites, \rev{while accounting for site- and year-level clustering}, we fitted generalized linear mixed models (GLMMs) to plot-level data. Because the experiment used a factorial $2 \times 2$ design, we modelled treatment effects as the planned factorial effects of Disturbance, NPK, and their interaction, rather than as a single four-level treatment factor. This parameterisation directly matches our biological questions: whether disturbance changes annual prevalence, whether nutrient addition changes annual prevalence, and whether \rev{disturbance and nutrient addition interact}. All models also included random intercepts for site and for year nested within site, \rev{capturing the key site and temporal variation relevant to estimating average network-wide treatment effects. We initially fitted more complex random structures, but these generally failed to converge, likely because the available within-site replication was insufficient to estimate the additional parameters reliably. We therefore retained this simpler structure. More details of the attempted models are provided in Appendix~\ref{appendixs1}.}

Different responses were matched to appropriate non-Gaussian distributions and link functions (for example, binomial models for species share, \rev{ordered} beta models for relative cover, negative-binomial models for richness, and zero-inflated lognormal models for absolute cover), chosen to respect the bounded or skewed nature of the data. 

For the main figures, we report treatment-wise estimated marginal means with 95\% confidence intervals. Estimated marginal means provide a model-based summary of \rev{the expected response under each treatment combination} for a ``typical'' site, defined as a site with random-effect values set to the network average on the model scale. These estimates therefore describe treatment differences after accounting for among-site and among-year variation, rather than simple averages that can be dominated by sites with extreme values or more observations.

For annual prevalence, estimated marginal means represent overall expected values on the response scale. For functional-group cover (Fig.~\ref{fig:functional_groups}), several groups were absent from most plots, especially annual graminoids, annual legumes and perennial legumes. In these cases, overall expected cover was strongly influenced by the high probability of zero cover (Tables~\ref{tab:table1_fig2}, \ref{tab:table5_fig2}, \ref{tab:table6_fig2}) and was therefore less informative about changes in cover where the group occurred. We therefore visualised the continuous component of the zero-inflated lognormal models for the functional-group analyses, representing expected cover when the focal group was present. In all cases, full model results, including zero-inflation components, are reported in Appendix~\ref{appendixs1}.

\medskip
\noindent
\textbf{Site-level variation in disturbance effects.}
Global averages do not reveal whether disturbance effects remain consistent across sites or vary with local context. We therefore estimated site-specific effects of Disturbance versus Control for annual species share and relative annual cover. To quantify proportional responses, we calculated log response ratios from site-level treatment means. To quantify absolute responses, we estimated additive differences in percentage points using simple models fitted separately within each site on the original response scale. We then tested whether site-level disturbance effects depended on initial community structure by relating log response ratios to baseline annual prevalence under Control using GAM smoothers.

\medskip
\noindent
\textbf{Climatic predictors and disturbance effects across climate space.}
To examine how annual prevalence and the impact of disturbance vary across climate, we aggregated plot-level data to site~$\times$~treatment means for annual species share and relative annual cover, and related these site-level proportions to climate. \rev{Because fertilisation did not consistently alter annual prevalence across the network, the climatic-envelope analysis focused on the contrasting Control and Disturbance treatments. Temperature and precipitation variables were extracted for each site from the WorldClim v2.1 bioclimatic layers at 30-arc-second spatial resolution, representing 1970--2000 climate normals.} We first compared \rev{ordered beta models} with different combinations of temperature and precipitation (or aridity) variables \rev{using the Control treatment only, with one observation per site,} and identified mean temperature of the warmest quarter and precipitation in the warmest quarter as the best pair of climatic predictors \rev{of baseline annual prevalence} (Table~\ref{tab:climate_long_table}).

We then modelled annual prevalence as a function of mean temperature and precipitation in the warmest quarter, \rev{Disturbance, and their interactions} in \rev{an ordered beta mixed-effects regression. Control and Disturbance observations were fitted jointly using the full factorial interaction between temperature, precipitation, and Disturbance, with a random intercept for site to account for the paired observations within sites. We compared this full model with a reduced model retaining the temperature~$\times$~precipitation interaction and the main effect of Disturbance, but excluding all Disturbance~$\times$~climate interactions. The reduced model had lower AIC, and the full interaction model did not significantly improve model fit in a likelihood-ratio test (Appendix Table~\ref{tab:climate_interaction_lrt}). We therefore generated predictions for Control and Disturbance from the reduced model across the observed climate space and visualised them with contour plots. 
We quantified the climatic envelope as the proportion of points on a regular prediction grid within the convex hull of observed site climates where predicted annual prevalence exceeded 50\%. These percentages describe climatic space rather than geographical area.} Model checks based on predictive simulations confirmed that the fitted climate \rev{model reproduced} the main gradients in our data \rev{(Appendix~\ref{appendixs1})}. Additional plots used to validate the figures in the paper, including \rev{the disturbance additive effect size metric and diagnostic simulations for the climate model}, can be found in Figures~\ref{fig:dist_additive}--\ref{fig:sim_unfert_spec}.

\section*{Acknowledgments}
This work was generated using data from the Disturbance and Recovery Across Grasslands experiment (DRAGNet), funded at the site-scale by individual researchers. Coordination and data management have been supported by funding to E. Borer and E. Seabloom from the National Science Foundation Research Coordination Network (NSF-DEB-2311608) and the Long Term Ecological Research program (DEB-1831944) to the Cedar Creek LTER. We also thank the UMN College of Biological Sciences for hosting project data.

\section*{Data availability}

All data and code required to reproduce the analyses and results reported in this manuscript are publicly available from the Environmental Data Initiative repository at \url{https://doi.org/10.6073/pasta/65799ea6ae60abdf0590b35134a9ab43}.

\bibliographystyle{vancouver}

\bibliography{references}  

\clearpage

\section{Authors and affiliations}

\author{%
  \parbox{\textwidth}{\raggedright
Niv DeMalach\textsuperscript{1}\orcidlink{0000-0002-4509-5387}\thanks{Corresponding author: \href{mailto:niv.demalach@mail.huji.ac.il}{\texttt{niv.demalach@mail.huji.ac.il}}}\allowbreak,
Roberto Salguero-Gomez\textsuperscript{2}\orcidlink{0000-0002-6085-4433},
Oded Hollander\textsuperscript{1},
Jane Lucas\textsuperscript{3}\orcidlink{0000-0002-3931-1864},
Alexander T. Strauss\textsuperscript{4}\orcidlink{0000-0003-0633-8443},

Tilak Chaudhary\textsuperscript{5}\orcidlink{0000-0002-8201-7142},
Y. Anny Chung\textsuperscript{6}\orcidlink{0000-0001-5207-2872},
James K. McCarthy\textsuperscript{7}\orcidlink{0000-0003-3060-1678},
Bruno Moreira\textsuperscript{8}\orcidlink{0000-0002-7319-2555},
Laís Petri\textsuperscript{9}\orcidlink{0000-0001-9727-1939},
Rachael Thornley\textsuperscript{2}\orcidlink{0000-0001-7651-0873},

Ingrid J. Slette\textsuperscript{10}\orcidlink{0000-0002-7707-3246},
Elizabeth T. Borer\textsuperscript{10}\orcidlink{0000-0003-2259-5853},
Eric W. Seabloom\textsuperscript{10}\orcidlink{0000-0001-6780-9259},

Joe Atkinson\textsuperscript{11}\orcidlink{0000-0001-9232-4421},
Karen H. Beard\textsuperscript{12}\orcidlink{0000-0003-4997-2495},
Lars A. Brudvig\textsuperscript{13}\orcidlink{0000-0002-3857-2165},
Christopher P. Catano\textsuperscript{14}\orcidlink{0000-0002-4060-7291},
Jane A. Catford\textsuperscript{15,16}\orcidlink{0000-0003-0582-5960},
Aimee Classen\textsuperscript{17}\orcidlink{0000-0002-6741-3470},
Adam Thomas Clark\textsuperscript{18}\orcidlink{0000-0002-8843-3278},
Ciara Dwyer\textsuperscript{19}\orcidlink{0000-0002-7558-3664},
Charles Fenster\textsuperscript{20}\orcidlink{0000-0002-1655-4409},
Oscar Godoy\textsuperscript{21}\orcidlink{0000-0003-4988-6626},
Or Gross\textsuperscript{1}\orcidlink{0009-0007-9202-7524},
Elizabeth Gusmán-Montalaván\textsuperscript{22}\orcidlink{0000-0002-3103-0419},
W. Stanley Harpole\textsuperscript{23,24,25}\orcidlink{0000-0002-3404-9174},
Erika Hersch-Green\textsuperscript{26}\orcidlink{0000-0003-3887-0768},
Petr Holub\textsuperscript{27}\orcidlink{0000-0002-3582-7900},
Anke Jentsch\textsuperscript{28}\orcidlink{0000-0002-2345-8300},
Gaurav Kandlikar\textsuperscript{29}\orcidlink{0000-0003-3043-6780},
Kevin Kirkman\textsuperscript{30}\orcidlink{0000-0001-9580-5191},
Sally E. Koerner\textsuperscript{31}\orcidlink{0000-0001-6403-7513},
Andrew Kulmatiski\textsuperscript{12}\orcidlink{0000-0001-9977-5508},
Jean-François Le Galliard\textsuperscript{32,33},
Daijiang Li\textsuperscript{34}\orcidlink{0000-0002-0925-3421},
Andrew S. MacDougall\textsuperscript{35},
Pierre Mariotte\textsuperscript{36}\orcidlink{0000-0001-8570-8742},
Jason P. Martina\textsuperscript{5}\orcidlink{0000-0002-3912-4911},
Pablo A. Meglioli\textsuperscript{37,38}\orcidlink{0000-0001-7296-3947},
Jonathan Millett\textsuperscript{39}\orcidlink{0000-0003-4701-3071},
Lydia O'Halloran\textsuperscript{40}\orcidlink{0000-0003-3634-9696},
Raúl Ochoa-Hueso\textsuperscript{41}\orcidlink{0000-0002-1839-6926},
Romà Ogaya\textsuperscript{42}\orcidlink{0000-0003-4927-8479},
Brooke Osborne\textsuperscript{43}\orcidlink{0000-0003-4771-7677},
Kalpana K. Patabendige\textsuperscript{44}\orcidlink{0000-0003-2249-2301},
Josep Peñuelas\textsuperscript{42,45}\orcidlink{0000-0002-7215-0150},
Pablo Peri\textsuperscript{46}\orcidlink{0000-0002-5398-4408},
Steven C. Pennings\textsuperscript{44}\orcidlink{0000-0003-4757-7125},
Anita C. Risch\textsuperscript{47}\orcidlink{0000-0003-0531-8336},
Christiane Roscher\textsuperscript{23,24}\orcidlink{0000-0001-9301-7909},
Nathan J. Sanders\textsuperscript{48}\orcidlink{0000-0001-6220-6731},
Carly Stevens\textsuperscript{49}\orcidlink{0000-0002-2390-1763},
Lauren Sullivan\textsuperscript{13,50}\orcidlink{0000-0002-4198-3483},
Pedro M. Tognetti\textsuperscript{51}\orcidlink{0000-0001-7358-1334},
Diego Vélez-Mora\textsuperscript{22}\orcidlink{0000-0003-1611-9842},
Cameron Wagg\textsuperscript{52}\orcidlink{0000-0002-9738-6901},
Chhaya Werner\textsuperscript{53}\orcidlink{0000-0002-2967-8603}
}}

\makeatletter
\renewcommand{\printaffiliations}{%
\noindent
\textsuperscript{1} Institute of Plant Sciences and Genetics in Agriculture, The Hebrew University of Jerusalem, Rehovot, Israel\\
\textsuperscript{2} Department of Biology, University of Oxford, Oxford OX1 3SZ, UK\\
\textsuperscript{3} Cary Institute of Ecosystem Studies\\
\textsuperscript{4} Odum School of Ecology, University of Georgia, Athens, GA 30606 USA\\
\textsuperscript{5} Department of Biology, Texas State University, San Marcos, TX 78666 USA\\
\textsuperscript{6} Departments of Plant Biology and Plant Pathology, University of Georgia, GA, USA\\
\textsuperscript{7} Bioeconomy Science Institute 7640, Lincoln, NZ\\
\textsuperscript{8} Centro de Investigaciones sobre Desertificación-CIDE (CSIC-UV-GVA), Valencia, Spain.\\
\textsuperscript{9} School of Biological Sciences, Southern Illinois University, Carbondale, IL, 62901, USA\\
\textsuperscript{10} Department of Ecology, Evolution, and Behavior, University of Minnesota, St Paul, MN 55108 USA\\
\textsuperscript{11} School of Biological Sciences, University of Adelaide, Adelaide, South Australia, Australia\\
\textsuperscript{12} Department of Wildland Resources and the Ecology Center, Utah State University, Logan, UT 84322 USA\\
\textsuperscript{13} Department of Plant Biology and Program in Ecology, Evolution, and Behavior, Michigan State University, East Lansing, MI 48824, USA\\
\textsuperscript{14} Department of Botany \& Plant Sciences, University of California Riverside, Riverside, CA 92521 USA\\
\textsuperscript{15} Department of Geography, King’s College London, 40 Aldwych, London, WC2B 4BG, UK\\
\textsuperscript{16} Fenner School of Environment \& Society, The Australian National University, Canberra, ACT 2600, Australia\\
\textsuperscript{17} Department of Ecology and Evolutionary Biology, University of Michigan, Ann Arbor, MI 48109; The University of Michigan Biological Station, University of Michigan, 9133 Biological Rd, Pellston, MI 49769\\
\textsuperscript{18} Department of Biology, University of Graz, Austria\\
\textsuperscript{19} Centre for Environment and Climate Science, Lund University, Sweden\\
\textsuperscript{20} Director, Oak Lake Field Station, South Dakota State University\\
\textsuperscript{21} Doñana Biological Station, Spanish Research Council (EBD-CSIC) Sevilla, Spain\\
\textsuperscript{22} Departamento de Ciencias Biológicas y Agropecuarias, Universidad Técnica Particular de Loja, Loja, Ecuador\\
\textsuperscript{23} Helmholtz Center for Environmental Research – UFZ, Department of Physiological Diversity, Permoserstrasse 15, 04318 Leipzig, Germany\\
\textsuperscript{24} German Centre for Integrative Biodiversity Research (iDiv), Puschstrasse 4, 04103 Leipzig, Germany\\
\textsuperscript{25} Martin Luther University Halle-Wittenberg, am Kirchtor 1, 06108 Halle (Saale), Germany\\
\textsuperscript{26} Department of Biological Sciences, Michigan Technological University, Houghton MI 49930\\
\textsuperscript{27} Global Change Research Institute of the Czech Academy of Sciences, Bělidla 4a, CZ-603 00 Brno, Czech Republic\\
\textsuperscript{28} Disturbance Ecology and Vegetation Dynamics, Bayreuth Center of Ecology and Environmental Studies (BayCEER), University of Bayreuth, Germany\\
\textsuperscript{29} Department of Biological Sciences, Louisiana State University, Baton Rouge, LA 70803 USA\\
\textsuperscript{30} School of Agriculture and Science, University of KwaZulu-Natal, Pietermaritzburg, South Africa\\
\textsuperscript{31} Department of Biology, University of North Carolina Greensboro, Greensboro, NC 27412 USA\\
\textsuperscript{32} Sorbonne Université, CNRS, IRD, INRAE, Institut d’écologie et des sciences de l’environnement (iEES Paris), 4 Place Jussieu, 75252 Paris Cedex 5, France\\
\textsuperscript{33} Ecole normale supérieure, PSL University, Département de biologie, CNRS, UMS 3194, Centre de recherche en écologie expérimentale et prédictive (CEREEP-Ecotron IleDeFrance), 11 chemin de Busseau, 77140 Saint-Pierre-lès-Nemours, France\\
\textsuperscript{34} Department of Biology, University of Wisconsin-Madison, Madison, WI, USA\\
\textsuperscript{35} College of Biological Science, Department of Integrative Biology, University of Guelph\\
\textsuperscript{36} Agroscope, Rte de la Tioleyre 4, 1725 Posieux, Switzerland\\
\textsuperscript{37} Instituto Argentino de Nivología, Glaciología y Ciencias Ambientales (IANIGLA), CONICET, Mendoza, Argentina\\
\textsuperscript{38} Facultad de Ciencias Agrarias (FCA), Universidad Nacional de Cuyo (UNCUYO), Luján de Cuyo, Mendoza, Argentina\\
\textsuperscript{39} Department of Geography \& Environment, Loughborough University, Loughborough, LE11 3TU, UK\\
\textsuperscript{40} Clemson University, Baruch Institute of Coastal Ecology and Forest Science\\
\textsuperscript{41} Department of Biology, IVAGRO, University of Cádiz, Campus de Excelencia Internacional Agroalimentario (CeiA3), Campus del Rio San Pedro, 11510 Puerto Real, Cádiz, Spain\\
\textsuperscript{42} CSIC, Global Ecology Unit CREAF–CSIC–UAB, Bellaterra, 08193 Barcelona, Catalonia, Spain\\
\textsuperscript{43} Department of Environment and Society, Utah State University, Moab, UT 84532, USA\\
\textsuperscript{44} Department of Biology and Biochemistry, University of Houston, Houston, TX 77204, USA\\
\textsuperscript{45} CREAF, Cerdanyola del Vallès, 08193 Barcelona, Catalonia, Spain\\
\textsuperscript{46} Instituto Nacional de Tecnología Agropecuaria (INTA)\\
\textsuperscript{47} Swiss Federal Institute for Forest, Snow and Landscape Research WSL, Zuercherstrasse 111, 8903 Birmensdorf, Switzerland\\
\textsuperscript{48} Department of Ecology and Evolutionary Biology, University of Michigan, Ann Arbor, MI 48109, USA\\
\textsuperscript{49} Lancaster Environment Centre, Lancaster University, Lancaster, LA1 4YQ, UK\\
\textsuperscript{50} W. K. Kellogg Biological Station, Michigan State University, Hickory Corners, MI 49060, USA\\
\textsuperscript{51} Universidad de Buenos Aires, Facultad de Agronomía; CONICET–IFEVA, Av. San Martín 4453, CABA, Argentina\\
\textsuperscript{52} Agriculture and Agri-Food Canada, Canada\\
\textsuperscript{53} Department of Environmental Science, Policy \& Sustainability, Southern Oregon University, Ashland, OR, USA\\
}  

\vspace{0.8\baselineskip} 

\printaffiliations
\newpage
\begin{longtable}{p{4.2cm}ccccccc}
\caption*{Data contribution statement (x = contribution). Analysis (lead) = analysed data; Analysis (contrib.) = contributed to data analysis; Writing (lead) = wrote the paper; Writing (contrib.) = contributed to paper writing; Site coord. = site coordinator; Net coord. = network-level coordinator.}\\
\toprule
Author & \shortstack{Analysis\\(lead)} & \shortstack{Analysis\\(contrib.)} & \shortstack{Writing\\(lead)} & \shortstack{Writing\\(contrib.)} & \shortstack{Site\\coord.} & \shortstack{Net\\coord.}\\
\midrule
\endfirsthead

\toprule
Author & \shortstack{Analysis\\(lead)} & \shortstack{Analysis\\(contrib.)} & \shortstack{Writing\\(lead)} & \shortstack{Writing\\(contrib.)} & \shortstack{Site\\coord.} & \shortstack{Net\\coord.}\\
\midrule
\endhead

\midrule
\multicolumn{7}{r}{\textit{continued on next page}}\\
\endfoot

\bottomrule
\endlastfoot

Niv DeMalach               & x &   & x &   & x &   \\
Roberto Salguero-Gomez     &   & x & x &   & x &   \\
Oded Hollander             & x &   &   & x &   &   \\
Jane Lucas                 & x &   &   & x & x &   \\
Alexander T. Strauss       & x &   &   & x & x &   \\

Tilak Chaudhary            &   & x &   & x &   &   \\
Y. Anny Chung              &   & x &   & x & x &   \\
James McCarthy             &   & x &   & x & x &   \\
Bruno Moreira              &   & x &   & x & x &   \\
Laís Petri                 &   & x &   & x &   &   \\
Rachael Thornley           &   & x &   & x &   &   \\

Ingrid J. Slette           &   &   &   & x &   & x \\
Elizabeth T. Borer         &   &   &   & x & x & x \\
Eric W. Seabloom           &   &   &   & x & x & x \\

Joe Atkinson               &   &   &   & x & x &   \\
Karen H. Beard             &   &   &   & x & x &   \\
Lars A. Brudvig            &   &   &   & x & x &   \\
Christopher P. Catano      &   &   &   & x & x &   \\
Jane A. Catford            &   &   &   & x & x &   \\
Aimee Classen              &   &   &   & x & x &   \\
Adam Thomas Clark          &   &   &   & x & x &   \\
Ciara Dwyer                &   &   &   & x & x &   \\
Charles Fenster            &   &   &   & x & x &   \\
Oscar Godoy                &   &   &   & x & x &   \\
Or Gross                   &   &   &   & x & x &   \\
Elizabeth Gusmán-Montalaván&   &   &   & x &   &   \\
W. Stanley Harpole         &   &   &   & x & x &   \\
Erika Hersch-Green         &   &   &   & x & x &   \\
Petr Holub                 &   &   &   & x & x &   \\
Anke Jentsch               &   &   &   & x & x &   \\
Gaurav Kandlikar           &   &   &   & x & x &   \\
Kevin Kirkman              &   &   &   & x & x &   \\
Sally E. Koerner           &   &   &   & x & x &   \\
Andrew Kulmatiski          &   &   &   & x & x &   \\
Jean-François Le Galliard  &   &   &   & x & x &   \\
Daijiang Li                &   &   &   & x & x &   \\
Andrew S. MacDougall       &   &   &   & x & x &   \\
Pierre Mariotte            &   &   &   & x & x &   \\
Jason P. Martina           &   &   &   & x & x &   \\
Pablo A. Meglioli          &   &   &   & x & x &   \\
Jonathan Millett           &   &   &   & x & x &   \\
Lydia O'Halloran           &   &   &   & x & x &   \\
Raúl Ochoa-Hueso           &   &   &   & x & x &   \\
Romà Ogaya                 &   &   &   & x & x &   \\
Brooke Osborne             &   &   &   & x & x &   \\
Kalpana K. Patabendige     &   &   &   & x & x &   \\
Josep Penuelas             &   &   &   & x & x &   \\
Pablo Peri                 &   &   &   & x & x &   \\
Steven C. Pennings         &   &   &   & x & x &   \\
Anita C. Risch             &   &   &   & x & x &   \\
Christiane Roscher         &   &   &   & x & x &   \\
Nathan J. Sanders          &   &   &   & x & x &   \\
Carly Stevens              &   &   &   & x & x &   \\
Lauren Sullivan            &   &   &   & x & x &   \\
Pedro M. Tognetti          &   &   &   & x & x &   \\
Diego Vélez-Mora           &   &   &   & x & x &   \\
Cameron Wagg               &   &   &   & x & x &   \\
Chhaya Werner              &   &   &   & x & x &   \\

\end{longtable}


\setcounter{section}{0}
\setcounter{figure}{0}
\setcounter{equation}{0}
\setcounter{table}{0}

\renewcommand{\theequation}{S\arabic{equation}}  
\renewcommand{\thesection}{S\arabic{section}}
\renewcommand{\thesubsection}{S\arabic{section}.\arabic{subsection}}
\renewcommand{\thefigure}{S\arabic{figure}}
\renewcommand{\thetable}{S\arabic{table}}
\section{Supporting Information}

\subsection*{Appendix S1. Details on the statistical modelling and its rationale}
\label{appendixs1}
 
Appendix~S1 provides additional details supporting the analyses reported in the main text. We first describe the data filtering and site inclusion criteria, then list the DRAGNet sites included in the study. We then provide the statistical details for the three main analytical components: network-level treatment effects, site-level variation in disturbance responses, and climate--disturbance models of annual prevalence. Because these analyses address different response variables and levels of aggregation, the sections below describe the model structure and rationale separately for each component.

\subsection*{Data cleaning and inclusion criteria}

We applied a small set of a priori filters to focus on sites and plots with data suitable for estimating treatment effects. First, we retained only sites with at least two years of observations, on the assumption that a single year is unreliable for annual plants with high year-to-year variability. Second, because disturbance was only applied during the early phase of the experiment, we restricted the analysis to the first three years after the start of the manipulation at each site; later years, in which disturbance ceased, were excluded. Third, we removed plot–year combinations with zero total vegetated cover, for which the proportion of annual cover and the proportion of annual species are undefined. 
\rev{Records classified as biennial (852 observations, ~3\% of the raw plot--species--year records prior to aggregation) were excluded because they do not map unambiguously onto the annual--perennial dichotomy that is the focus of this study.}
After these filters, the dataset comprised 37 sites and 1,444 plot–year observations.

\clearpage
\subsection*{Study Sites} 

\begin{longtable}{llll}
\caption{Description of the 37 DRAGNet sites used in this study.} \label{table:sites} \\
  \hline
  \textbf{Site Name} & \textbf{Site Code} & \textbf{Country} & \textbf{Coordinates} \\
  \hline
  \endfirsthead
  \hline
  \textbf{Site Name} & \textbf{Site Code} & \textbf{Country} & \textbf{Coordinates} \\
  \hline
  \endhead
  \hline
  \endfoot
  Ainsdale Dune Slacks            & ains\_dn.uk      & UK & 53.59°N, 3.07°W   \\
  Agroscope Changins              & chan\_dn.ch      & CH & 46.40°N, 6.23°E   \\
  Algaida                         & alga\_dn.es      & ES & 36.54°N, 6.20°W   \\
  Archbold Biological Station     & arch\_dn.us      & US & 27.17°N, 81.22°W  \\
  Bad Lauchstaedt                 & badlau\_dn.de    & DE & 51.38°N, 11.88°E  \\
  Bayreuth DRAGNet                & bayr\_dn.de      & DE & 49.92°N, 11.58°E  \\
  Bladen Lakes State Forest       & blsf\_dn.us      & US & 34.72°N, 78.49°W  \\
  Caracoles                       & cara\_dn.es      & ES & 37.07°N, 6.32°W   \\
  Cary IES                        & cary\_dn.us      & US & 41.79°N, 73.74°W  \\
  Cass                            & cass\_dn.nz      & NZ & 43.04°S, 171.75°E \\
  Cedar Creek LTER                & cedr\_dn.us      & US & 45.40°N, 93.20°W  \\
  CEREEP -- Ecotron IDF           & cereep\_dn.fr    & FR & 48.28°N, 2.67°E   \\
  Churning Rapids                 & chur\_dn.us      & US & 47.18°N, 88.61°W  \\
  Cowichan                        & cowi\_dn.ca      & CA & 48.81°N, 123.63°W \\
  Dauphin Island                  & daup\_dn.us      & US & 30.25°N, 88.20°W  \\
  Domaninek                       & doma\_dn.cz      & CZ & 49.54°N, 16.25°E  \\
  Hazelrigg                       & lancaster\_dn.uk & UK & 54.01°N, 2.78°W   \\
  Jena                            & jena\_dn.de      & DE & 50.94°N, 11.53°E  \\
  Kellogg Biological Station      & kell\_dn.us      & US & 42.42°N, 85.37°W  \\
  Kniezenberg                     & kzbg\_dn.at      & AT & 46.96°N, 15.27°E  \\
  Meranges                        & mera\_dn.es      & ES & 42.43°N, 1.78°E   \\
  Millville                       & mill\_dn.us      & US & 41.65°N, 111.81°W \\
  Nillahcootie                    & nilla\_dn.au     & AU & 36.90°S, 146.01°E \\
  Park Shaked                     & shak\_dn.il      & IL & 31.27°N, 34.65°E  \\
  Piedmont Prairie                & pied\_dn.us      & US & 33.89°N, 83.36°W  \\
  Potrok Aike                     & potrok\_dn.ar    & AR & 51.92°S, 70.41°W  \\
  San Claudio                     & sanc\_dn.ar      & AR & 35.92°S, 61.16°W  \\
  Sandflats                       & sdft\_dn.us      & US & 38.58°N, 109.49°W \\
  Sierra Foothills REC            & sier\_dn.us      & US & 39.26°N, 121.32°W \\
  Shortgrass Steppe LTER          & sgs\_dn.us       & US & 40.82°N, 104.77°W \\
  Temple                          & temple\_dn.us    & US & 31.10°N, 97.34°W  \\
  Ukulinga                        & ukul\_dn.za      & ZA & 29.67°S, 30.40°E  \\
  Univ.\ of Michigan Bio.\ Station & umbs\_dn.us     & US & 45.56°N, 84.68°W  \\
  Wallula Gap -- North Face       & wgno\_dn.us      & US & 46.00°N, 118.91°W \\
  Wallula Gap -- South Face       & wgso\_dn.us      & US & 46.00°N, 118.95°W \\
  WSL                             & wald\_dn.ch      & CH & 47.36°N, 8.46°E   \\
  Wytham Woods                    & wyth\_dn.uk      & UK & 51.77°N, 1.33°W   \\         
  \hline
  \end{longtable}
\subsection*{(i) Network-level treatment effects}

To estimate the overall effects of disturbance and fertilisation across the DRAGNet network, we fitted generalized linear mixed models (GLMMs) to plot-level data using the \texttt{glmmTMB} package. This framework allows us to combine non-Gaussian distributions (beta, negative binomial, lognormal), zero-inflation components, and hierarchical random effects suitable for the multi-site, repeated-measures design.

For each response, we modelled the expected value at plot $i$ in site $j$ and year $k$ as
\[
g(\mu_{ijk}) \;=\; \mathbf{X}_{ijk}\boldsymbol{\beta} \;+\; b_{j} \;+\; u_{jk},
\]
where $\mathbf{X}_{ijk}$ contains the fixed effects of the Disturbance $\times$ NPK factorial treatment (Disturbance and NPK coded as 0/1), $\boldsymbol{\beta}$ are fixed-effect coefficients, $b_{j}$ is a random intercept for site, and $u_{jk}$ is an additional random intercept for year nested within site. Here $g(\cdot)$ is the link function that maps the mean response $\mu_{ijk}$ to the linear predictor; the specific choice of link and response family depends on the outcome and is described below for each analysis.

\rev{The retained models included random intercepts for site and for year nested within site. The site intercept accounts for the large baseline differences among sites, while the year-within-site intercept accounts for shared year-to-year variation within sites. \rev{These random effects capture major among-site and year-to-year variation relevant to estimating the network-wide treatment effects that are the focus of this study.} Accordingly, the fixed treatment coefficients represent average network-wide treatment effects during the active treatment period.}

\rev{We initially explored more detailed random-effect structures, including random intercepts for blocks nested within sites, site-level random slopes for Disturbance and NPK, and plot-level random intercepts to account explicitly for repeated measurements of the same plots. In almost all response models, these additional structures failed to converge reliably. This is not surprising given the relatively limited amount of data within each site, with only 3--5 blocks and two to three active-treatment years. We therefore retained the simpler structure described above. Consequently, the models account for baseline differences among sites and year-to-year variation within sites, but do not estimate variation among blocks, site-specific treatment effects, or the additional repeated-measure dependence at the individual-plot level.}

Year was treated as a random effect nested within site rather than as a fixed factor. \rev{This allows us to account for year-to-year variability without estimating a common temporal trend.} Calendar years are not aligned across sites because treatment initiation occurred in different years, so a global fixed ``year'' effect would be difficult to interpret ecologically. \rev{Moreover, sites contributed only two to three active-treatment years. Because disturbance was applied annually, observations from later years reflect responses following repeated disturbance rather than to a single isolated event. Temporal trends are nevertheless difficult to estimate reliably over such short time series, because apparent temporal change combines directional responses with year-to-year variability, and both are likely to differ among sites. We therefore did not estimate a common temporal trajectory or the effect of each successive disturbance event. For the two primary response variables, variance associated with year within site was substantially smaller than among-site variance (Tables~\ref{tab:zibetaregression}--\ref{tab:logisregression}), suggesting that overall year-to-year variability was relatively modest compared with differences among sites.}

Different responses required different distributions. \rev{All mixed-effects models described below were fitted using the \texttt{glmmTMB} package, allowing us to retain a common fixed- and random-effects structure while matching each response to an appropriate distribution.} Relative annual cover (proportion of total vegetated cover) is a proportion on $[0, 1]$ with about $30\%$ exact zeros and \rev{8\% exact ones}. We modelled this response with \rev{an ordered beta mixed-effects regression with a logit link}. \rev{Ordered beta regression extends beta regression to the closed interval $[0,1]$, allowing exact zeros and ones to be retained without transformation or separate zero- and one-inflation components. The model assumes that boundary and intermediate values represent different positions along a single underlying continuous process. A common linear predictor describes this underlying process; two estimated cutpoints determine the probabilities of observations occurring at exactly 0 or 1, while observations between these boundaries follow a beta distribution. This formulation is particularly appropriate for relative annual cover because values from 0 to 1 naturally describe a continuum from entirely perennial to entirely annual vegetation, rather than requiring absence, intermediate cover, and complete annual dominance to arise from separate ecological processes.}

Annual species share (annual richness divided by total richness) is naturally expressed as a binomial outcome: ``number of annual species out of total species''. We therefore used a binomial GLMM with a logit link, treating annual richness as the number of successes and total richness as the number of trials. Absolute richness of annuals and perennials is overdispersed count data, so we used negative-binomial GLMMs with a log link. Absolute cover of annuals and perennials, both at the whole-community level and within each functional group (graminoids, forbs, legumes), can be zero and can exceed $100\%$ when overlapping canopies are summed. These responses are continuous, right-skewed, and contain frequent zeros, so we modelled them with zero-inflated lognormal GLMMs (with a log link), which handle the zeros and positive values separately. 

Treatment effects in the main figures are reported as estimated marginal means (with their corresponding CIs) from the fitted GLMMs, computed with the \texttt{emmeans} package. These estimates are model-based expectations for each treatment combination, evaluated at a ``typical'' site (random effects set to zero) on the linear predictor scale and back-transformed to the response scale. In a heterogeneous, unbalanced multi-site network, raw averages of plot-level data can be misleading because they are strongly influenced by sites with extreme values or many plots. Marginal means provide a more interpretable summary of the typical treatment \rev{effect} across the network, because they respect the link function and distributional assumptions and are not dominated by a small number of highly variable sites. For bounded indices such as proportions, averaging on the linear predictor (e.g. logit) scale and back-transforming typically yields values closer to the median than the simple arithmetic mean, which again aligns with the idea of a typical site.

\rev{For the ordered beta models,  \texttt{emmeans} returns only $\mu$, the location parameter of the model's continuous component, which does not incorporate the probability mass at the boundaries ($Y=0$, $Y=1$) and can therefore differ materially from the true expected response when boundary mass is appreciable, as it is here. We therefore computed the full expected response, $E[Y] = P(Y{=}1) + P(0{<}Y{<}1)\cdot\mu$, directly from the fitted coefficients and the two estimated cutpoints, and obtained its confidence interval by bootstrapping: simulating 5,000 draws from the joint asymptotic covariance of the fixed-effect parameters (including the cutpoints) and taking the 2.5th and 97.5th percentiles of the resulting distribution.}

Model adequacy for these GLMMs was checked using simulation-based diagnostics from the \texttt{DHARMa} package. We inspected quantile--quantile plots, residuals versus fitted values, and tests for dispersion and zero inflation, and did not detect systematic deviations that would compromise inference about the \rev{fixed treatment effects}.

\subsection*{(ii) Site-level comparison of effect size}

The GLMMs described above quantify overall \rev{treatment effects} across the network. To investigate how the effect of disturbance varies among individual sites, we also conducted site-specific analyses that focus on two complementary effect size metrics: the log response ratio and the additive difference in percentage points.

\begin{figure}[H]
\centering
\includegraphics[width=0.8\columnwidth]{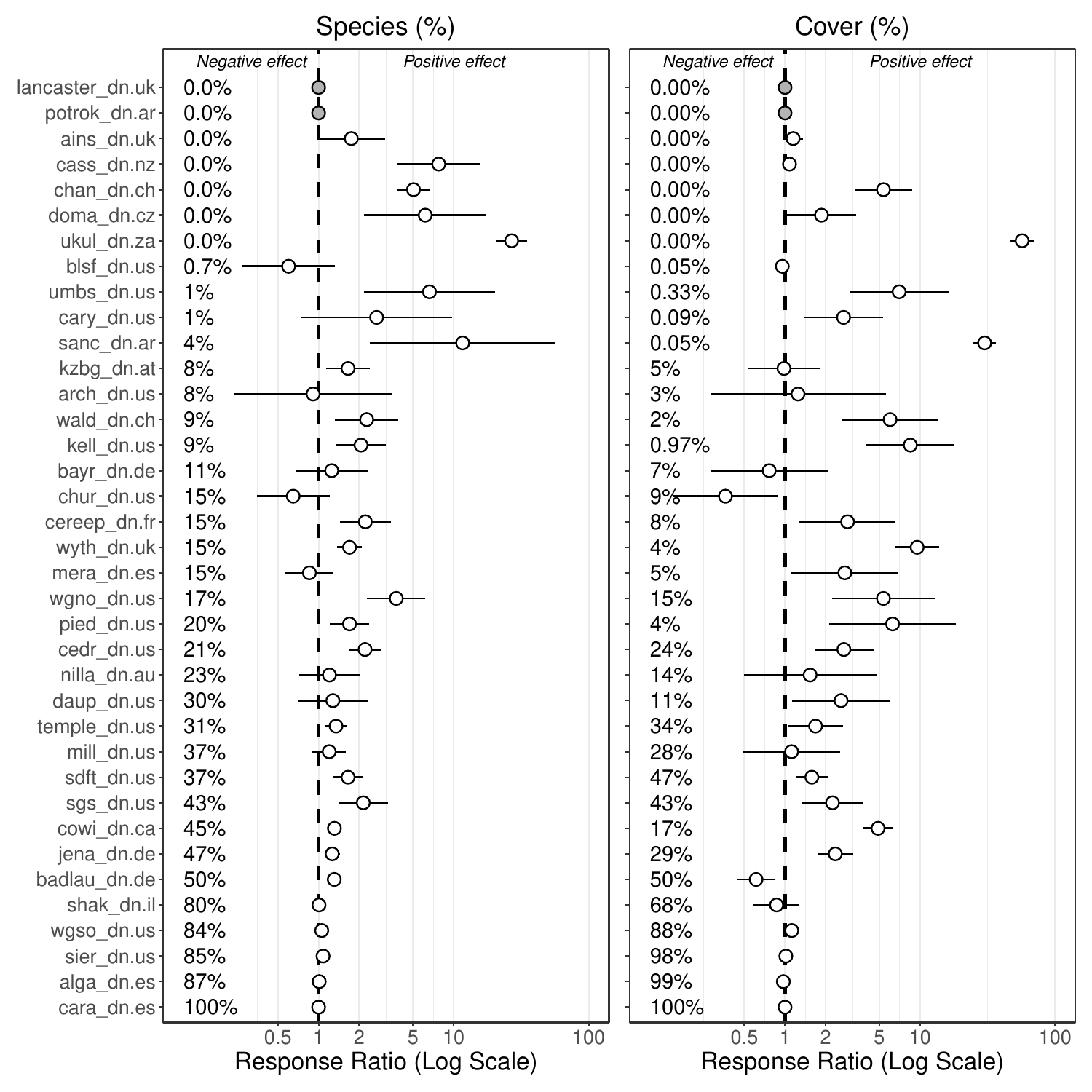}
\caption{%
\textbf{Disturbance drives increases in annual prevalence across most sites.}
  \rev{The left panel shows site-level disturbance effects based on the average annual species share across plots, while the right panel shows corresponding effects on relative annual cover}. Axes show disturbance effects as response ratios, where
  values above 1 indicate higher annual prevalence in disturbed plots and values below 1 indicate lower annual prevalence. For statistical calculation, effect sizes were computed on the log scale as
  log(Disturbance / Control) and then back-transformed for display. Sites are ordered (top to bottom) starting with sites with zero annual prevalence in both control and treatment plots (marked with grey
  circles), followed by the remaining sites ordered by baseline annual species share. Percent values shown beside each row indicate the baseline annual value for the response displayed in that panel:
  annual species share in the left panel and relative annual cover in the right panel. Uncertainty is shown by the 95\% confidence intervals. See table~\ref{table:sites} for additional site information.
  }
\label{fig:dist_multi}
\end{figure}

For the \emph{log response ratio} (Fig.~\ref{fig:dist_multi}), we worked with site-level summaries. For each site, the response (e.g.\ relative annual cover) was averaged across plots and years within each treatment (Control and Disturbance) to obtain site-level means $\bar{y}_C$ and $\bar{y}_D$. Before computing the log ratio, a constant of 1\% was added to all means to yield the adjusted log response ratio $\log[(\bar{y}_D+1)/(\bar{y}_C+1)]$. This procedure ensures the ratio is well-defined at sites where $\bar{y}=0$, preventing the exclusion of sites where annuals were absent. Additionally, the constant serves to stabilize the delta-method standard error approximation, which involves the squared mean in the denominator and would otherwise become poorly behaved as $\bar{y}$ approaches zero. The same adjusted means are utilized in the SE formula:

\[
\mathrm{SE}\biggl[\log\bigg(\frac{\bar{y}_D + 1}{\bar{y}_C + 1}\bigg)\biggr]
\;\approx\;
\sqrt{
  \frac{s_D^2}{n_D (\bar{y}_D + 1)^2}
  +
  \frac{s_C^2}{n_C (\bar{y}_C + 1)^2}
},
\]

assuming independence between treatments. Given that the constant is small relative to typical values (median control baseline $\approx 7\%$ for cover and $\approx 7\%$ for species share), its influence is negligible for sites with substantial annual presence while ensuring all sites contribute to the site-level comparison. Repeating the analysis without adding the constant for the subset of sites with non-zero means in both treatments yielded the same overall qualitative pattern and very similar effect sizes for nearly all sites. The main discrepancies occurred only at sites with mean annual cover close to zero, where unadjusted log response ratios became extremely large and were likely driven by boundary sensitivity rather than biologically meaningful differences.

\begin{figure}[H]
    \centering
    \includegraphics[width=0.8\linewidth]{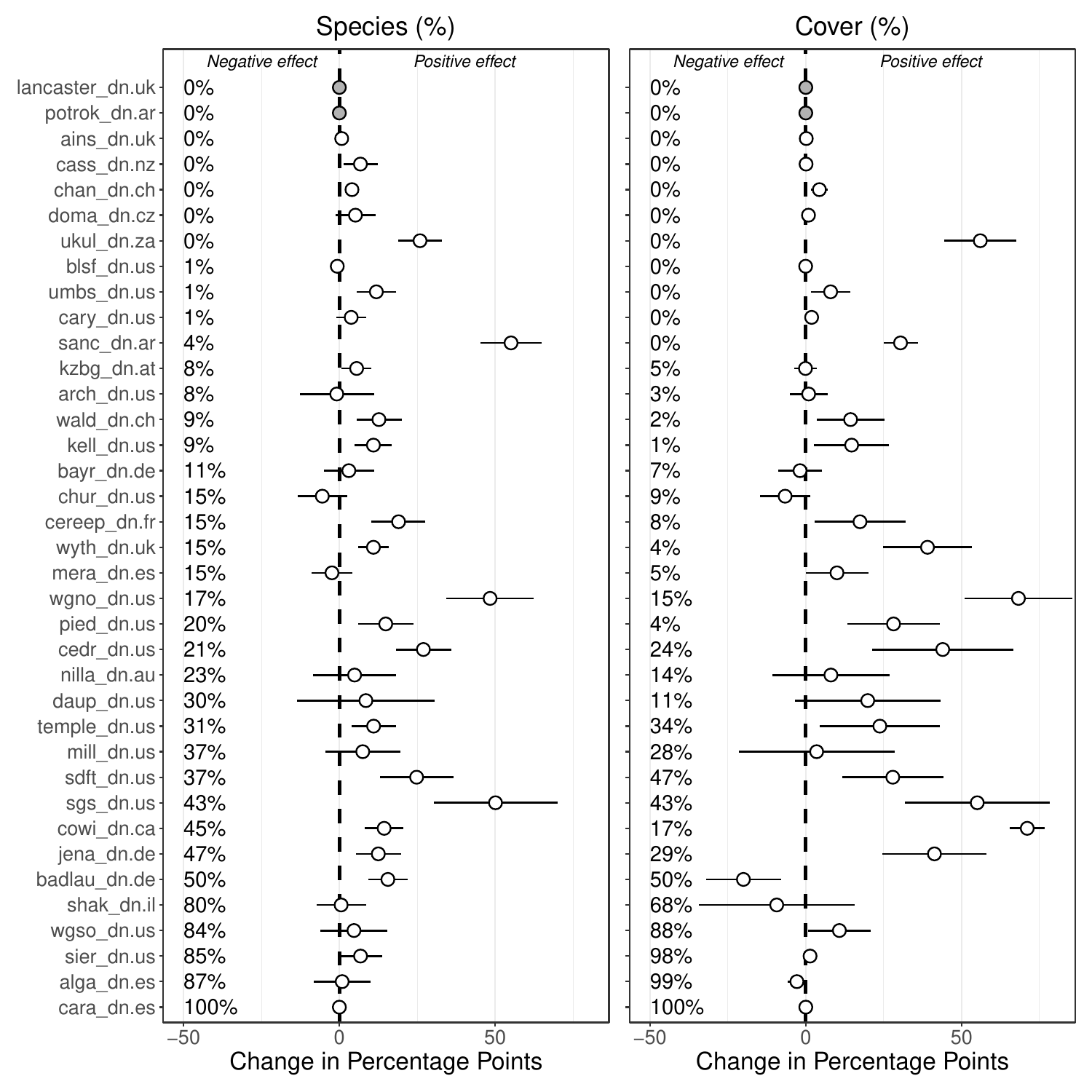}
    \caption{%
\textbf{Disturbance drives absolute increases in the prevalence of annuals}
      \rev{The left panel shows site-level disturbance effects based on the average annual species share across plots, while the right panel shows corresponding effects on relative annual cover.} Effect sizes were calculated as the difference in
      treatment means ($\mathrm{Disturbance} - \mathrm{Control}$). Sites are ordered (top to bottom) starting with sites that had zero annuals in both control and treatment plots (marked with grey
     circles), followed by the remaining sites ordered by baseline annual species share. Horizontal error bars represent 95\% confidence intervals. Percent values shown beside each row indicate the
     baseline annual value for the response displayed in that panel: annual species share in the left panel and relative annual cover in the right panel. See table~\ref{table:sites} for additional site
     information.
     }
    \label{fig:dist_additive}
\end{figure}

\begin{figure}
    \centering
    \includegraphics[width=0.8\linewidth]{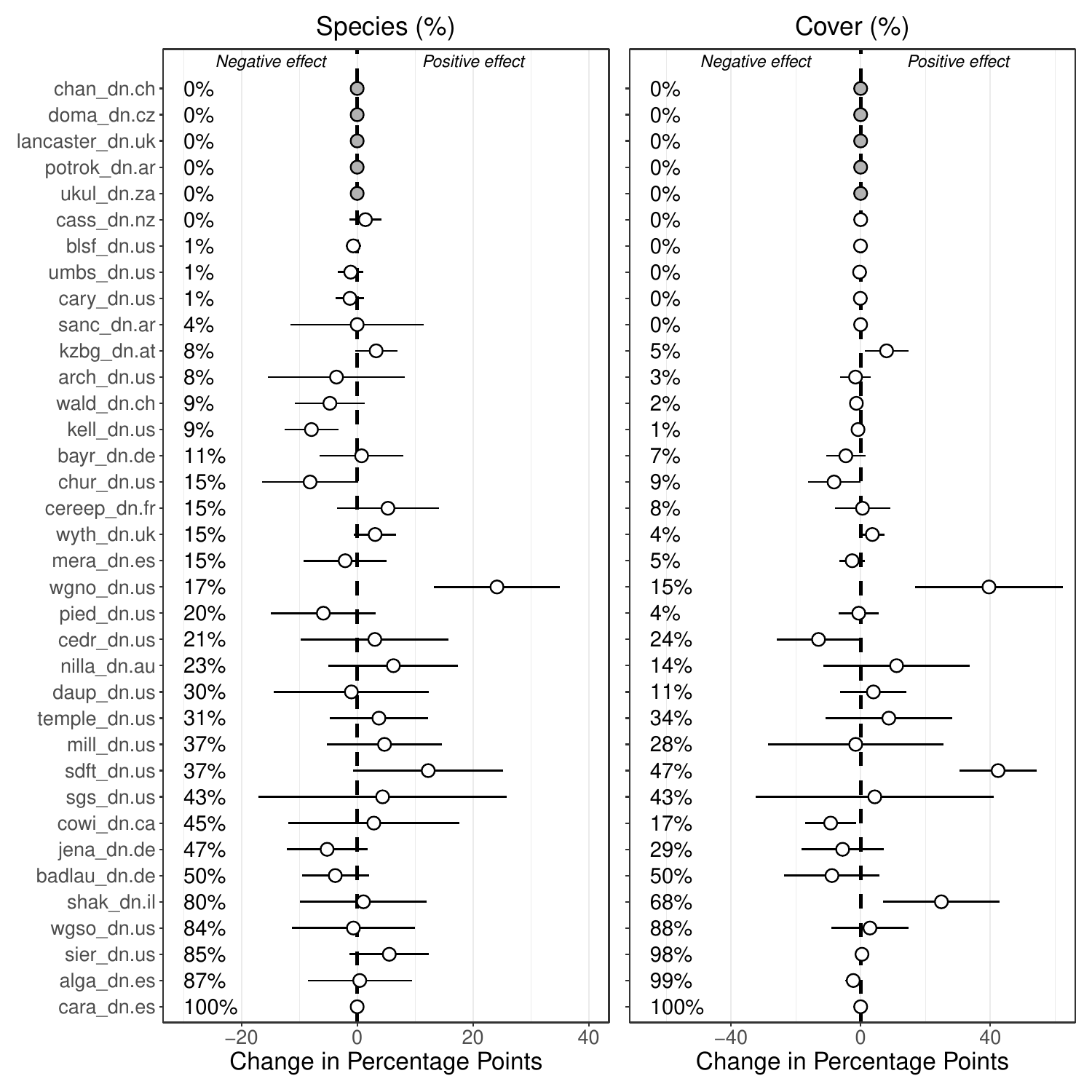}
    \caption{%
\textbf{NPK addition results in mixed absolute effects on annual prevalence, with substantial increases in cover observed at specific sites.}
     The panels display the absolute change in annual species share (left) and relative annual cover (right). Effect sizes represent the difference in means ($\mathrm{NPK} - \mathrm{Control}$). Horizontal
      error bars show 95\% confidence intervals calculated using robust standard errors. Sites are ordered (top to bottom) starting with sites that had zero annuals in both control and treatment plots
     (marked with grey circles), followed by the remaining sites ordered by baseline annual species share. One site (ains dn.uk) was excluded from this analysis due to missing data for the NPK treatment.
      Percent values shown beside each row indicate the baseline annual value for the response displayed in that panel: annual species share in the left panel and relative annual cover in the right panel.
      See table~\ref{table:sites} for additional site information.
     }
    \label{fig:npk_additive}
\end{figure}

For the \emph{additive effect} (Figs.~\ref{fig:dist_additive}, \ref{fig:npk_additive}), we returned to the plot-level data. The goal here was to quantify the absolute change in percentage points between Disturbance and Control on the original response scale. By definition, such an additive \rev{effect} must be defined on the identity (response) scale. \rev{Models with non-identity link functions, such as logit or log links, are additive on the link scale and therefore do not directly estimate an additive difference in percentage points on the response scale}. We therefore estimated the additive effect using simple linear regression fitted separately within each site to the plot-level observations. In this formulation, the treatment coefficient corresponds directly to the additive treatment difference in percentage points for that site. To account for heteroscedasticity, we obtained heteroscedasticity-robust standard errors for these site-specific linear models using the \texttt{sandwich} package. Sites with zero variance (e.g., 0\% annuals across all plots and treatments) were assigned an additive effect of zero with a standard error of zero for display in the forest plots; this reflects the absence of observed variation at these sites and should not be interpreted as an inferential estimate of perfect precision.

The resulting log response ratios and additive effects, with their respective uncertainties, were displayed in forest plots to enable direct comparison of proportional and absolute disturbance effects across the network.
To visualise how disturbance effects depend on baseline community structure, we related site-level effect sizes to the baseline annual proportion (under Control) using generalised additive models fitted with \texttt{mgcv}, allowing a smooth, data-driven relationship without imposing a specific functional form.

\subsection*{(iii) Climate \rev{envelope} and combined climate--disturbance models}

The climate analyses address how annual prevalence varies across global climate gradients and how disturbance alters the climatic space of annual dominance. Because climate covariates such as temperature and precipitation are defined at the site level, we first aggregated the plot-level data to site $\times$ treatment means, so that each site contributes a single value per treatment for \rev{relative annual cover and annual species share, each calculated at the plot level and then averaged across plots and years}. After aggregation, these responses are continuous proportions on $[0, 1]$ that can take the values 0 and 1 exactly.

We first standardised all candidate climate variables by $z$-scoring within the network to improve numerical stability and make coefficients comparable. Our goal was to identify a pair of climate axes, one thermal and one moisture-related, that best explain variation in \rev{baseline} annual prevalence. Following the idea that plant performance is jointly constrained by temperature and water availability, but without imposing \emph{a priori} assumptions about which specific metrics matter most, we considered a broad set of temperature and precipitation variables, including annual means, ranges, and seasonal (quarter-based) indices. We also included the aridity index (precipitation / potential evapotranspiration). Although aridity is affected by temperature, we treated it as a ``corrected precipitation'' metric within our moisture-axis set. Monthly variables were not included because they are highly collinear with the quarter-based variables and harder to interpret ecologically. A complete list of candidate climate variables is given in Table~\ref{tab:climate_long_table}.

For each response (relative annual cover and species share), we then fitted \rev{ordered beta regressions using only the Control treatment, with one observation per site}. \rev{Models used a logit link and a constant beta precision parameter.} We fitted all combinations of one temperature variable and one precipitation (or aridity) variable, \rev{including their interaction in the linear predictor}. For each combination we calculated the Akaike Information Criterion (AIC). \rev{Because only one Control observation was included per site, no site-level random effect was required in this screening stage.} This structured search reflects the hypothesis that \rev{baseline} annual prevalence is primarily associated with one temperature axis and one water-availability axis, while avoiding over-fitting and interpretational difficulties that would arise from including many correlated covariates in the same model. For both responses, mean temperature of the warmest quarter and precipitation of the warmest quarter produced the lowest AIC. For relative annual cover, no alternative combination was within $\Delta$AIC $< 2$, whereas for annual species share two alternative temperature metrics combined with precipitation of the warmest quarter were within $\Delta$AIC $< 2$ (Table~\ref{tab:climate_long_table}). We therefore retained mean temperature and precipitation of the warmest quarter as the common climatic predictors for the subsequent analyses and visualisations. 

\rev{Having selected these two climatic predictors, we next asked whether the effect of disturbance varied across climate. For this analysis, we fitted Control and Disturbance observations jointly using an ordered beta mixed-effects model with mean temperature of the warmest quarter ($T_{\mathrm{warmQ}}$), precipitation in the warmest quarter ($P_{\mathrm{warmQ}}$), disturbance, and their interactions as predictors. We first fitted the full factorial interaction $T_{\mathrm{warmQ}} \times P_{\mathrm{warmQ}} \times D$, including all corresponding main effects and lower-order interactions. A random intercept for site was included to account for the paired Control and Disturbance observations within sites. We then compared this model with a reduced model retaining the $T_{\mathrm{warmQ}} \times P_{\mathrm{warmQ}}$ interaction and the main effect of disturbance, but excluding all disturbance~$\times$~climate interactions. Model comparison provided no support for retaining the disturbance~$\times$~climate interactions (Table~\ref{tab:climate_interaction_lrt}), so subsequent predictions and visualisations were based on the reduced model.}

\rev{We generated predicted annual prevalence from the reduced model on a regular grid covering the convex hull of the observed mean temperature in the warmest quarter $\times$ precipitation in the warmest quarter points, thereby avoiding extrapolation beyond the sampled climate space. As in the network-wide ordered beta analysis, these predictions represent the full expected response, including the probability mass at $Y=0$ and $Y=1$. Predictions were generated separately for Control and Disturbance, with the site random intercept set to its mean of zero.} Contour plots of \rev{predicted annual prevalence} were used to visualise the climatic surface of annual prevalence, and we highlighted regions where \rev{predicted annual prevalence} exceeds $0.5$, defining the climatic ``dominance zone'' of annuals in our network. \rev{To visualise how the absolute disturbance effect varies across climate space, we mapped the difference between predicted annual prevalence under Disturbance and Control, expressed in percentage points.}

\begin{figure}
    \centering
    \includegraphics[width=1\linewidth]{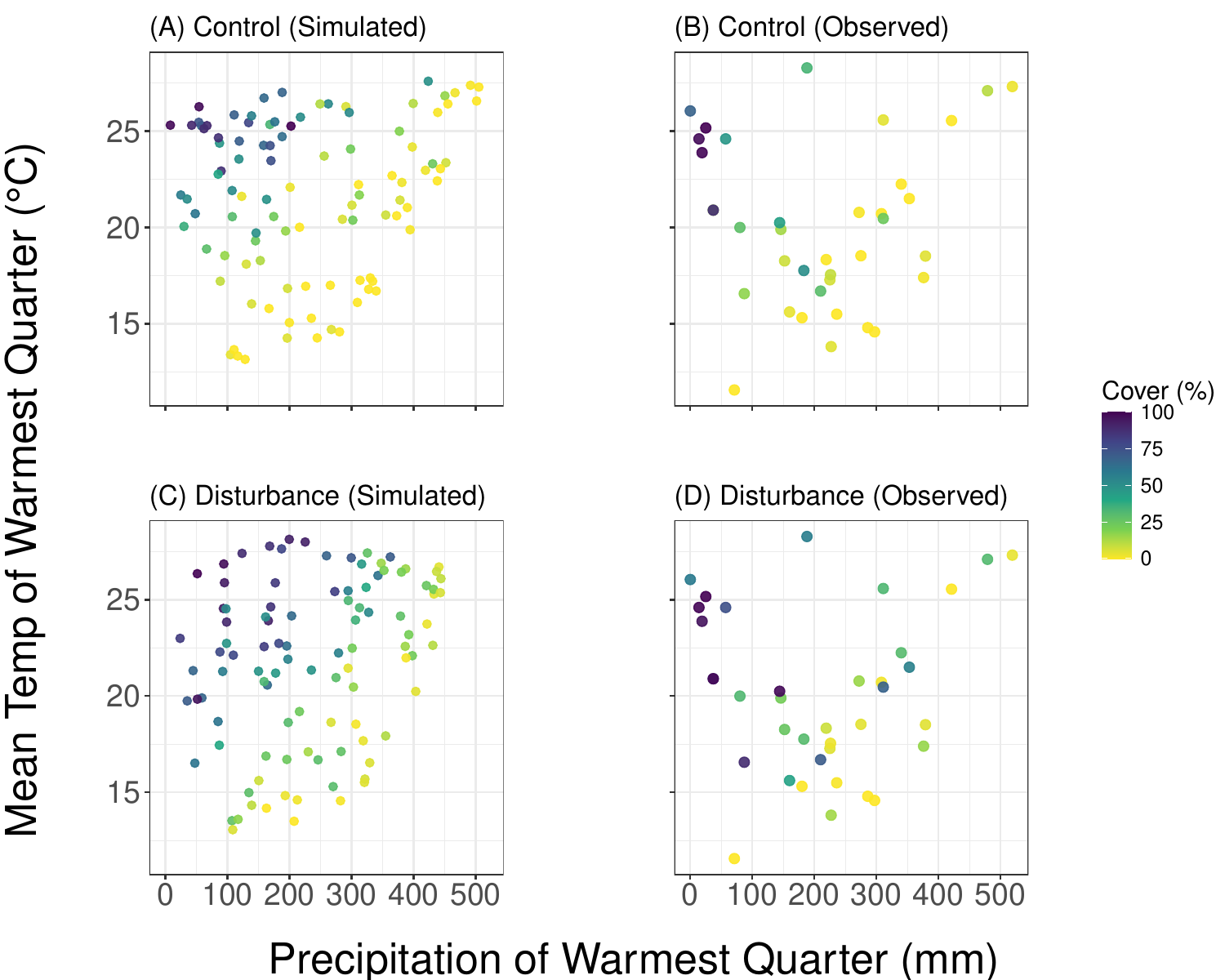}
    \caption{
\textbf{The fitted models accurately reproduce the observed distribution of annual cover (unfertilised plots).}
Comparison of simulated (left column) and observed (right column) annual plant cover across the climate space defined by mean temperature and precipitation of the warmest quarter.
\rev{(A, C) Simulated datasets generated by drawing random climate locations uniformly within the convex hull of the study sites and sampling from the fitted ordered beta models.}
(B, D) Empirical field observations from the control and disturbance treatments.
The color scale indicates the percentage of total vegetation cover comprised of annuals. The structural similarity between the simulated predictive draws and the observed data demonstrates that the models adequately capture the climatic gradients and the distribution of proportion values (including zeros and ones).
}
    \label{fig:sim_unfert_cover}
\end{figure}

\begin{figure}
    \centering
    \includegraphics[width=1\linewidth]{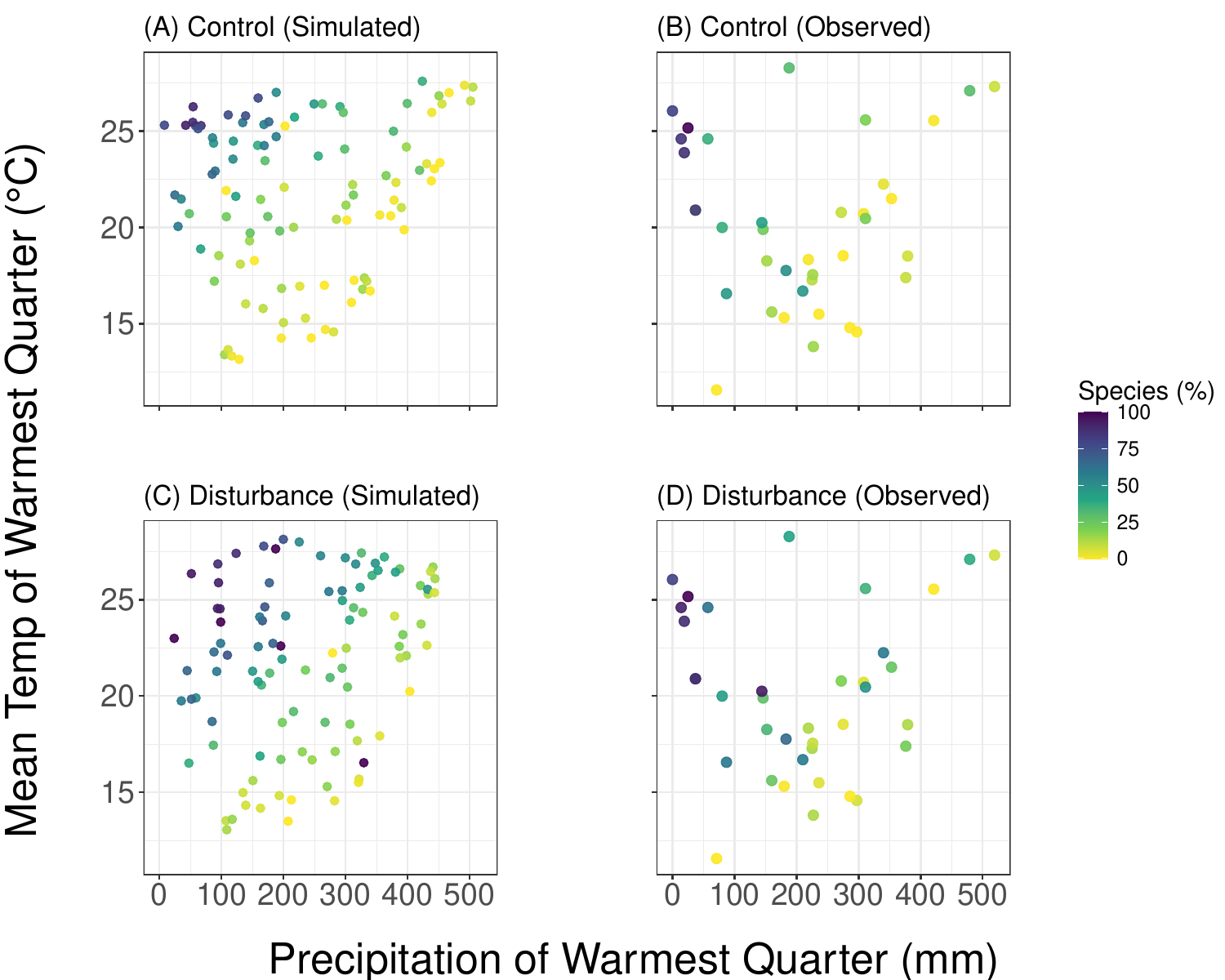}
    \caption{
\textbf{The fitted models accurately reproduce the observed distribution of annual species share (unfertilised plots).}
Comparison of simulated (left column) and observed (right column) annual species richness proportions across the climate space.
\rev{(A, C) Simulated datasets generated from the fitted ordered beta models using random climate draws within the climatic hull of the network.}
(B, D) Observed field data showing the proportion of total species richness identified as annuals at each site.
Colors represent the percentage of annual species, from 0\% (yellow) to 100\% (dark purple).
These diagnostics confirm that the fitted models successfully reproduce the observed patterns of species dominance, including the response to temperature and precipitation gradients under both control and disturbance conditions.
}
    \label{fig:sim_unfert_spec}
\end{figure}

\rev{We additionally used predictive simulations to evaluate whether the fitted climate model could reproduce the observed data across climate space. The climatic response surfaces are intended to represent the fitted climate--treatment relationship itself, so the site random intercept was set to its mean of zero when generating these predictions. Predictive model checks serve a different purpose: they ask whether observations generated from the fitted mixed-effects model resemble the actual observations, including their among-site variability and the occurrence of values at the boundaries of 0 and 1. We therefore sampled climate points within the same convex hull used for the response surfaces and drew site random intercepts from the fitted site-level distribution. For each simulated site, we generated paired Control and Disturbance responses from the same reduced ordered beta model, using the same climate values and site random effect for both treatments and changing only the Disturbance indicator. We then compared the simulated and observed site-level responses across the same climate space. This comparison provides a check that the fitted model captures not only the broad climatic gradients in mean annual prevalence, but also the observed spread among sites and the frequency of boundary values. The simulated data reproduced these main features of the observations (Figs.~\ref{fig:sim_unfert_cover}--\ref{fig:sim_unfert_spec}), providing additional support that the response surfaces are representative of the underlying data rather than artifacts of the fitted parametric form.}

\begin{figure}[H]
    \centering    \includegraphics[width=0.8\linewidth]{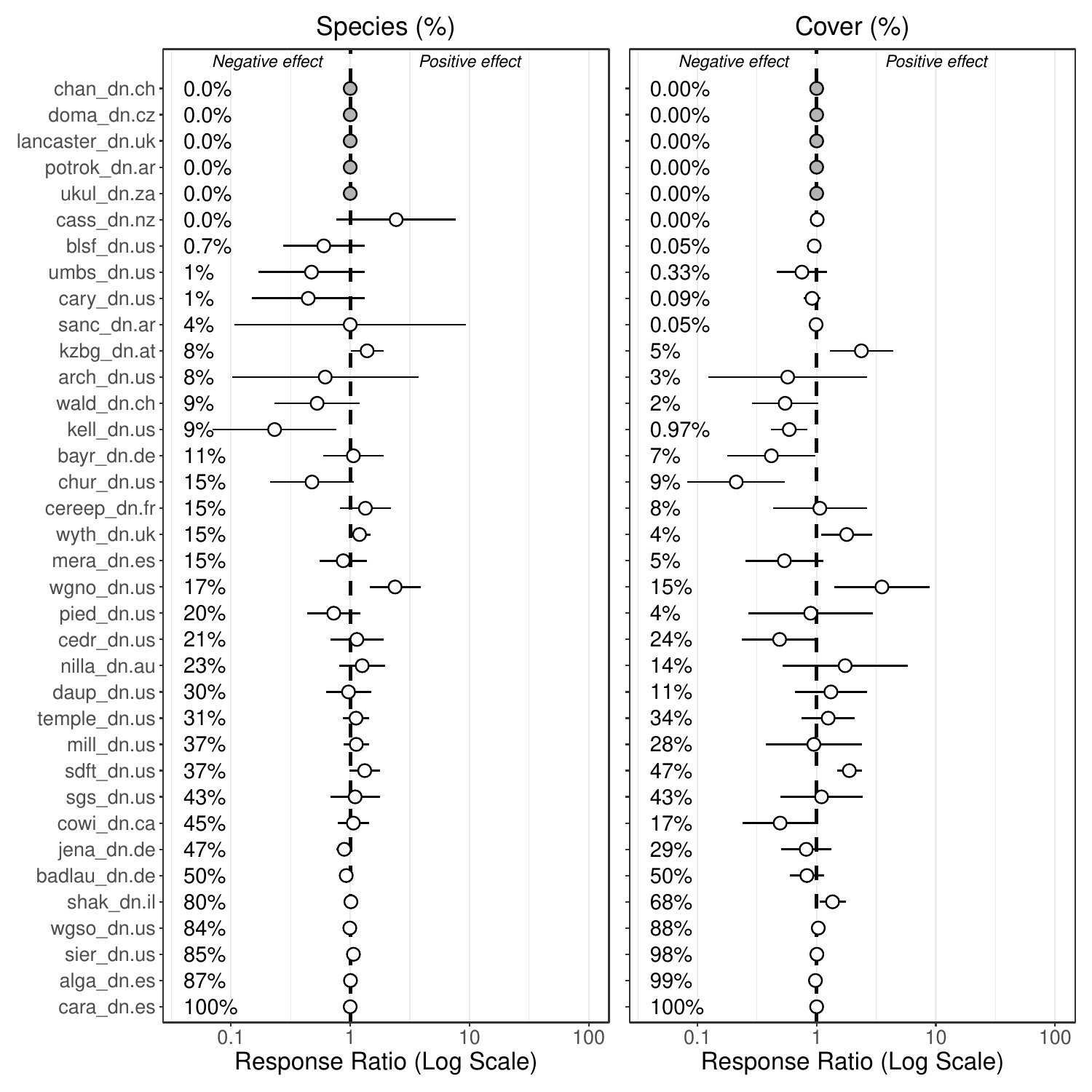}
\caption{%
     \textbf{Unlike disturbance, NPK addition elicits variable responses in annual prevalence, with no consistent directional trend.}  \rev{The left panel shows site-level fertilization effects based on the average annual species share across plots, while the right panel shows corresponding effects on relative annual cover.}
     Effect sizes were calculated as log response ratios; axes tick labels represent back-transformed response ratios. Horizontal error bars denote 95\% confidence intervals. Sites are ordered (top to
     bottom) starting with sites that had zero annuals in both control and treatment plots (marked with grey circles), followed by the remaining sites ordered by baseline annual species share. Percent
     values shown beside each row indicate the baseline annual value for the response displayed in that panel: annual species share in the left panel and relative annual cover in the right panel. One site
      (ains dn.uk) was excluded from this analysis due to missing data for the NPK treatment. See table~\ref{table:sites} for additional site information.
     }
    \label{fig:npk_multi}
\end{figure}

\begin{table}[H]
\footnotesize
\caption{Ordered beta GLMM for relative annual cover.
The ordered beta model treats relative annual cover, including exact values of 0 and 1, as outcomes along a single underlying bounded process.
The \textit{Mu (Mean)} component describes treatment effects on the model's logit-linked annual-cover scale.
The intercept describes the reference treatment (Control: no disturbance, no NPK), and treatment coefficients describe shifts relative to this reference treatment.
Because coefficients in the Mu component are exponentiated, values greater than 1 indicate a shift toward higher relative annual cover and values below 1 indicate a shift toward lower relative annual cover.
\textit{Disturbance} and \textit{NPK} test the main treatment effects, and \textit{Disturbance} $\times$ \textit{NPK} tests their interaction.
\textit{Sigma (Precision)} describes the precision of intermediate proportions.
\textit{Cutpoints (Boundary)} gives the estimated lower and upper cutpoints governing the probabilities of observations at exactly 0 and 1.
Random-effect terms report variance components for site and site-by-year grouping structure; ICC is the intraclass correlation.
$N_{\text{site}} = 37$.}
\label{tab:zibetaregression}
\centering
\begin{tabular}{lccc}
\multicolumn{4}{c}{\textbf{Relative annual cover}} \\[4pt]
\hline
\textit{Predictors} & \textit{Estimates} & \textit{CI} & \textit{p} \\
\hline\\[-12pt]

\textbf{Mu (Mean)} & & & \\[2pt]
\quad (Intercept)              & 0.14 & 0.06 -- 0.35 & \textbf{$<$0.001} \\[2pt]
\quad Disturbance              & 3.22 & 2.72 -- 3.82 & \textbf{$<$0.001} \\[2pt]
\quad NPK                      & 1.17 & 0.99 -- 1.38 & 0.058 \\[2pt]
\quad Disturbance $\times$ NPK & 1.08 & 0.87 -- 1.36 & 0.481 \\[2pt]

\\[-12pt]
\textbf{Sigma (Precision)} & & & \\[2pt]
\quad (Intercept)              & 1.47    & 1.38 -- 1.57       & \textbf{$<$0.001} \\[4pt]

\\[-12pt]
\textbf{Cutpoints (Boundary)} & & & \\[2pt]
\quad Lower cutoff             & $-$2.91 & $-$3.11 -- $-$2.71 & \textbf{$<$0.001} \\[2pt]
\quad Upper cutoff             & 3.23    & 2.95 -- 3.51       & \textbf{$<$0.001} \\[2pt]

\\[-6pt]
\textbf{Random Effects} & & & \\[4pt]
\quad $\sigma^{2}$                                  & 0.44 & & \\[4pt]
\quad $\tau_{00}$ \scriptsize{site\_name}           & 7.42 & & \\[4pt]
\quad $\tau_{00}$ \scriptsize{site\_name:year\_trt} & 0.25 & & \\[4pt]
\quad ICC                                           & 0.95 & & \\[4pt]
\quad N \scriptsize{site\_name}                     & 37   & & \\[4pt]
\quad N \scriptsize{year\_trt}                      & 3    & & \\[4pt]
\quad Observations                                  & 1444 & & \\[4pt]
\quad Marginal R\textsuperscript{2} / Conditional R\textsuperscript{2} & 0.045 / 0.948 & & \\[4pt]
\hline
\end{tabular}
\end{table}

\clearpage

\begin{table}[H]
    \caption{Binomial GLMM for annual species share, calculated as annual richness divided by total richness per plot.
The intercept describes the reference treatment (Control: no disturbance, no NPK) and represents the baseline odds of a species being annual.
Treatment coefficients describe multiplicative effects on these odds relative to the reference treatment. Values greater than 1 indicate a higher annual species share, whereas values below 1 indicate a lower annual species share.
\textit{Disturbance} and \textit{NPK} test the main treatment effects, and \textit{Disturbance} $\times$ \textit{NPK} tests their interaction.
Random-effect terms report variance components for site and site-by-year treatment grouping structure; ICC is the intraclass correlation. $N_{\text{site}} = 37$.}

    \label{tab:logisregression}
    \centering
    \begin{tabular}{lccc}
        \multicolumn{4}{c}{\textbf{proportion of annual species}} \\[4pt]
    \hline
    \textit{Predictors} & \textit{Estimates} & \textit{CI} & \textit{p} \\
    \hline\\[-6pt]
    (Intercept)              & 0.18 & 0.09 -- 0.39 & \textbf{$<$0.001} \\[4pt]
    Disturbance              & 2.20 & 1.93 -- 2.52 & \textbf{$<$0.001} \\[4pt]
    NPK                      & 1.13 & 0.98 -- 1.30 & 0.104 \\[4pt]
    Disturbance $\times$ NPK & 0.91 & 0.75 -- 1.10 & 0.335 \\[4pt]
    \\[-6pt]
    \textbf{Random Effects} & & & \\[4pt]
    \quad $\sigma^{2}$                                   & 0.37 & & \\[4pt]
    \quad $\tau_{00}$ \scriptsize{site\_name}            & 5.28 & & \\[4pt]
    \quad $\tau_{00}$ \scriptsize{site\_name:year\_trt}  & 0.03 & & \\[4pt]
    \quad ICC                                            & 0.94 & & \\[4pt]
    \quad N \scriptsize{site\_name}                      & 37   & & \\[4pt]
    \quad N \scriptsize{year\_trt}                       & 3    & & \\[4pt]
    \quad Observations                                  & 1444 & & \\[4pt]
    \hline
    \end{tabular}
    \end{table}


\begin{table}[H]
    \caption{Negative binomial GLMM for annual species richness, calculated as the number of annual species per plot.
    The intercept describes the reference treatment (Control: no disturbance, no NPK) and represents the baseline expected annual species richness.
    Treatment coefficients describe multiplicative effects relative to this reference treatment. Values greater than 1 indicate higher annual species richness, whereas values below 1 indicate lower annual species richness.
    The negative binomial family was used to accommodate overdispersion in count data, as supported by DHARMa diagnostics.
    \textit{Disturbance} and \textit{NPK} test the main effects; \textit{Disturbance\,$\times$\,NPK} tests their interaction.
    Random Effects ($\tau_{00}$) report between-site and between-year variance; ICC is the intraclass correlation. $N_{\text{site}} = 37$.}
    \label{tab:neg_binomial_annual}
    \centering
    \begin{tabular}{lccc}
        \multicolumn{4}{c}{\textbf{annual species richness}} \\[4pt]
    \hline
    \textit{Predictors} & \textit{Estimates} & \textit{CI} & \textit{p} \\
    \hline\\[-6pt]
    (Intercept)              & 1.10 & 0.68 -- 1.80 & 0.693 \\[4pt]
    Disturbance              & 1.46 & 1.34 -- 1.59 & \textbf{$<$0.001} \\[4pt]
    NPK                      & 0.91 & 0.83 -- 1.01 & 0.065 \\[4pt]
    Disturbance $\times$ NPK & 1.05 & 0.92 -- 1.18 & 0.479 \\[4pt]
    \\[-6pt]
    \textbf{Random Effects} & & & \\[4pt]
    \quad $\sigma^{2}$                                   & 0.22 & & \\[4pt]
    \quad $\tau_{00}$ \scriptsize{site\_name}            & 2.19 & & \\[4pt]
    \quad $\tau_{00}$ \scriptsize{site\_name:year\_trt}  & 0.04 & & \\[4pt]
    \quad ICC                                            & 0.91 & & \\[4pt]
    \quad N \scriptsize{site\_name}                      & 37   & & \\[4pt]
    \quad N \scriptsize{year\_trt}                       & 3    & & \\[4pt]
    \quad Observations                                  & 1444 & & \\[4pt]
    \hline
    \end{tabular}
    \end{table}
  

\begin{table}[H]
    \caption{Negative binomial GLMM for perennial species richness, calculated as the number of perennial species per plot.
    Model structure and interpretation follow Table~\ref{tab:neg_binomial_annual}.
    The intercept represents the baseline expected perennial species richness in the reference treatment (Control: no disturbance, no NPK).
    Treatment coefficients describe multiplicative effects relative to this reference treatment. Values greater than 1 indicate higher perennial species richness, whereas values below 1 indicate lower perennial species richness.}
    \label{tab:neg_binomial_perenn}
    \centering
    \begin{tabular}{lccc}
        \multicolumn{4}{c}{\textbf{perennial species richness}} \\[4pt]
    \hline
    \textit{Predictors} & \textit{Estimates} & \textit{CI} & \textit{p} \\
    \hline\\[-6pt]
    (Intercept)              & 5.31 & 3.84 -- 7.34 & \textbf{$<$0.001} \\[4pt]
    Disturbance              & 0.81 & 0.76 -- 0.85 & \textbf{$<$0.001} \\[4pt]
    NPK                      & 0.81 & 0.76 -- 0.85 & \textbf{$<$0.001} \\[4pt]
    Disturbance $\times$ NPK & 1.18 & 1.09 -- 1.29 & \textbf{$<$0.001} \\[4pt]
    \\[-6pt]
    \textbf{Random Effects} & & & \\[4pt]
    \quad $\sigma^{2}$                                   & 0.13 & & \\[4pt]
    \quad $\tau_{00}$ \scriptsize{site\_name}            & 0.98 & & \\[4pt]
    \quad $\tau_{00}$ \scriptsize{site\_name:year\_trt}  & 0.03 & & \\[4pt]
    \quad ICC                                            & 0.89 & & \\[4pt]
    \quad N \scriptsize{site\_name}                      & 37   & & \\[4pt]
    \quad N \scriptsize{year\_trt}                       & 3    & & \\[4pt]
    \hline
    Observations & 1444 & & \\
    \hline
    \end{tabular}
    \end{table}
  

\begin{table}[H]
\footnotesize
    \caption{Zero-inflated lognormal GLMM for absolute annual cover.
    The \textit{Continuous Model} describes annual cover among plots where annual cover is present, whereas the \textit{Zero-Inflated Model} describes the probability of zero annual cover. In each model component, the intercept describes the reference treatment (Control: no disturbance, no NPK): the baseline geometric mean annual cover among non-zero plots in the Continuous Model, and the baseline odds of zero annual cover in the Zero-Inflated Model.
    Treatment coefficients describe multiplicative effects relative to this reference treatment. In the Continuous Model, values greater than 1 indicate higher annual cover and values below 1 indicate lower annual cover. In the Zero-Inflated Model, values greater than 1 indicate a higher probability of zero annual cover and values below 1 indicate a lower probability of zero annual cover.
    \textit{Disturbance} and \textit{NPK} test the main effects; \textit{Disturbance\,$\times$\,NPK} tests their interaction.
  $N_{\text{site}} = 37$.}
    \label{tab:zi_lognormal_annual}
    \centering
    \begin{tabular}{lccc}
        \multicolumn{4}{c}{\textbf{annual cover}} \\[4pt]
    \hline
    \textit{Predictors} & \textit{Estimates} & \textit{CI} & \textit{p} \\
    \hline\\[-6pt]
    \textbf{Continuous Model} & & & \\[4pt]
    \quad (Intercept)              & 20.20 & 15.42 -- 26.46 & \textbf{$<$0.001} \\[4pt]
    \quad Disturbance              & 1.35  & 1.21 -- 1.51   & \textbf{$<$0.001} \\[4pt]
    \quad NPK                      & 1.17  & 1.05 -- 1.30   & \textbf{0.004} \\[4pt]
    \quad Disturbance $\times$ NPK & 1.01  & 0.89 -- 1.16   & 0.856 \\[4pt]
    \\[-6pt]
    \textbf{Zero-Inflated Model} & & & \\[4pt]
    \quad (Intercept)              & 0.21  & 0.05 -- 0.91   & \textbf{0.037} \\[4pt]
    \quad Disturbance              & 0.11  & 0.06 -- 0.20   & \textbf{$<$0.001} \\[4pt]
    \quad NPK                      & 1.34  & 0.80 -- 2.25   & 0.265 \\[4pt]
    \quad Disturbance $\times$ NPK & 0.71  & 0.33 -- 1.53   & 0.381 \\[4pt]
    \\[-6pt]
    \textbf{Random Effects} & & & \\[4pt]
    \quad $\sigma^{2}$                                   & 43.72 & & \\[4pt]
    \quad $\tau_{00}$ \scriptsize{site\_name}            & 0.51  & & \\[4pt]
    \quad $\tau_{00}$ \scriptsize{site\_name:year\_trt}  & 0.12  & & \\[4pt]
    \quad ICC                                            & 0.01  & & \\[4pt]
    \quad N \scriptsize{site\_name}                      & 37    & & \\[4pt]
    \quad N \scriptsize{year\_trt}                       & 3     & & \\[4pt]
   \quad Observations                                  & 1444 & & \\[4pt]
    \hline
    \end{tabular}
    \end{table}


\begin{table}[H]
    \caption{Zero-inflated lognormal GLMM for absolute perennial cover.
    Model structure and interpretation follow Table~\ref{tab:zi_lognormal_annual}.
    }
    \label{tab:zi_lognormal_perenn}
    \centering
    \begin{tabular}{lccc}
        \multicolumn{4}{c}{\textbf{perennial cover}} \\[4pt]
    \hline
    \textit{Predictors} & \textit{Estimates} & \textit{CI} & \textit{p} \\
    \hline\\[-6pt]
    \textbf{Continuous Model} & & & \\[4pt]
    \quad (Intercept)              & 66.48 & 51.63 -- 85.60 & \textbf{$<$0.001} \\[4pt]
    \quad Disturbance              & 0.59  & 0.55 -- 0.63   & \textbf{$<$0.001} \\[4pt]
    \quad NPK                      & 1.03  & 0.98 -- 1.08   & 0.219 \\[4pt]
    \quad Disturbance $\times$ NPK & 1.13  & 1.04 -- 1.23   & \textbf{0.003} \\[4pt]
    \\[-6pt]
    \textbf{Zero-Inflated Model} & & & \\[4pt]
    \quad (Intercept)              & 0.00  & 0.00 -- 0.00   & \textbf{$<$0.001} \\[4pt]
    \quad Disturbance              & 2.96  & 1.31 -- 6.70   & \textbf{0.009} \\[4pt]
    \quad NPK                      & 1.22  & 0.51 -- 2.90   & 0.657 \\[4pt]
    \quad Disturbance $\times$ NPK & 1.59  & 0.52 -- 4.87   & 0.415 \\[4pt]
    \\[-6pt]
    \textbf{Random Effects} & & & \\[4pt]
    \quad $\sigma^{2}$                                   & 35.27 & & \\[4pt]
    \quad $\tau_{00}$ \scriptsize{site\_name}            & 0.57  & & \\[4pt]
    \quad $\tau_{00}$ \scriptsize{site\_name:year\_trt}  & 0.03  & & \\[4pt]
    \quad ICC                                            & 0.02  & & \\[4pt]
    \quad N \scriptsize{site\_name}                      & 37    & & \\[4pt]
    \quad N \scriptsize{year\_trt}                       & 3     & & \\[4pt]
    \quad Observations                                  & 1444 & & \\[4pt]
    \hline
    \end{tabular}
    \end{table}


  \begin{table}[H]
    \caption{Zero-inflated lognormal GLMM for absolute annual graminoid cover (\%).
    Model structure and interpretation follow Table~\ref{tab:zi_lognormal_annual}.
    }
    \label{tab:table1_fig2}
    \centering
    \begin{tabular}{lccc}
    \multicolumn{4}{c}{\textbf{Annual graminoid cover (\%)}} \\[4pt]
    \hline
    \textit{Predictors} & \textit{Estimates} & \textit{CI} & \textit{p} \\
    \hline\\[-6pt]
        \textbf{Continuous Model} & & & \\[4pt]
    \quad (Intercept)              & 15.10 & 11.01 -- 20.71  & \textbf{$<$0.001} \\[4pt]
    \quad Disturbance              & 1.09  & 0.92 -- 1.29    & 0.314 \\[4pt]
    \quad NPK                      & 1.65  & 1.41 -- 1.92    & \textbf{$<$0.001} \\[4pt]
    \quad Disturbance $\times$ NPK & 0.75  & 0.61 -- 0.92    & \textbf{0.005} \\[4pt]
    \\[-6pt]
    \textbf{Zero-Inflated Model} & & & \\[4pt]
    \quad (Intercept)              & 53.35 & 2.02 -- 1406.36 & \textbf{0.017} \\[4pt]
    \quad Disturbance              & 0.55  & 0.29 -- 1.06    & 0.073 \\[4pt]
    \quad NPK                      & 1.05  & 0.55 -- 1.99    & 0.881 \\[4pt]
    \quad Disturbance $\times$ NPK & 0.68  & 0.27 -- 1.69    & 0.407 \\[4pt]
    \\[-6pt]
    \textbf{Random Effects} & & & \\[4pt]
    \quad $\sigma^{2}$                                   & 39.68 & & \\[4pt]
    \quad $\tau_{00}$ \scriptsize{site\_name}            & 0.28  & & \\[4pt]
    \quad $\tau_{00}$ \scriptsize{site\_name:year\_trt}  & 0.16  & & \\[4pt]
    \quad ICC                                            & 0.01  & & \\[4pt]
    \quad N \scriptsize{site\_name}                      & 37    & & \\[4pt]
    \quad N \scriptsize{year\_trt}                       & 3     & & \\[4pt]
    \quad Observations                                  & 1444 & & \\[4pt]
    \hline
    \end{tabular}
    \end{table}


  \begin{table}[H]
    \caption{Zero-inflated lognormal GLMM for absolute perennial graminoid cover (\%).
   Model structure and interpretation follow Table~\ref{tab:zi_lognormal_annual}.}
    \label{tab:table2_fig2}
    \centering
    \begin{tabular}{lccc}
    \multicolumn{4}{c}{\textbf{Perennial graminoid cover (\%)}} \\[4pt]
    \hline
    \textit{Predictors} & \textit{Estimates} & \textit{CI} & \textit{p} \\
    \hline\\[-6pt]
    \textbf{Continuous Model} & & & \\[4pt]
    \quad (Intercept)              & 43.69 & 35.18 -- 54.26 & \textbf{$<$0.001} \\[4pt]
    \quad Disturbance              & 0.51  & 0.47 -- 0.56   & \textbf{$<$0.001} \\[4pt]
    \quad NPK                      & 1.09  & 1.02 -- 1.16   & \textbf{0.012} \\[4pt]
    \quad Disturbance $\times$ NPK & 1.15  & 1.02 -- 1.29   & \textbf{0.019} \\[4pt]
    \\[-6pt]
    \textbf{Zero-Inflated Model} & & & \\[4pt]
    \quad (Intercept)              & 0.01  & 0.00 -- 0.10   & \textbf{$<$0.001} \\[4pt]
    \quad Disturbance              & 2.45  & 1.36 -- 4.41   & \textbf{0.003} \\[4pt]
    \quad NPK                      & 0.86  & 0.45 -- 1.64   & 0.655 \\[4pt]
    \quad Disturbance $\times$ NPK & 1.57  & 0.68 -- 3.65   & 0.290 \\[4pt]
    \\[-6pt]
    \textbf{Random Effects} & & & \\[4pt]
    \quad $\sigma^{2}$                                   & 35.84 & & \\[4pt]
    \quad $\tau_{00}$ \scriptsize{site\_name}            & 0.35  & & \\[4pt]
    \quad $\tau_{00}$ \scriptsize{site\_name:year\_trt}  & 0.05  & & \\[4pt]
    \quad ICC                                            & 0.01  & & \\[4pt]
    \quad N \scriptsize{site\_name}                      & 37    & & \\[4pt]
    \quad N \scriptsize{year\_trt}                       & 3     & & \\[4pt]
    \quad Observations                                  & 1444 & & \\[4pt]
    \hline
    \end{tabular}
    \end{table}
  

  \begin{table}[H]
    \caption{Zero-inflated lognormal GLMM for absolute annual forb cover (\%).
    Model structure and interpretation follow Table~\ref{tab:zi_lognormal_annual}.}
    \label{tab:table3_fig2}
    \centering
    \begin{tabular}{lccc}
    \multicolumn{4}{c}{\textbf{Annual forb cover (\%)}} \\[4pt]
    \hline
    \textit{Predictors} & \textit{Estimates} & \textit{CI} & \textit{p} \\
    \hline\\[-6pt]
        \textbf{Continuous Model} & & & \\[4pt]
    \quad (Intercept)              & 12.29 & 9.49 -- 15.91 & \textbf{$<$0.001} \\[4pt]
    \quad Disturbance              & 1.72  & 1.52 -- 1.96  & \textbf{$<$0.001} \\[4pt]
    \quad NPK                      & 1.00  & 0.87 -- 1.15  & 0.962 \\[4pt]
    \quad Disturbance $\times$ NPK & 1.08  & 0.91 -- 1.27  & 0.384 \\[4pt]
    \\[-6pt]
    \textbf{Zero-Inflated Model} & & & \\[4pt]
    \quad (Intercept)              & 0.81  & 0.23 -- 2.85  & 0.742 \\[4pt]
    \quad Disturbance              & 0.11  & 0.07 -- 0.19  & \textbf{$<$0.001} \\[4pt]
    \quad NPK                      & 1.45  & 0.91 -- 2.31  & 0.115 \\[4pt]
    \quad Disturbance $\times$ NPK & 0.99  & 0.50 -- 1.95  & 0.978 \\[4pt]
    \\[-6pt]
    \textbf{Random Effects} & & & \\[4pt]
    \quad $\sigma^{2}$                                   & 36.90 & & \\[4pt]
    \quad $\tau_{00}$ \scriptsize{site\_name}            & 0.39  & & \\[4pt]
    \quad $\tau_{00}$ \scriptsize{site\_name:year\_trt}  & 0.08  & & \\[4pt]
    \quad ICC                                            & 0.01  & & \\[4pt]
    \quad N \scriptsize{site\_name}                      & 37    & & \\[4pt]
    \quad N \scriptsize{year\_trt}                       & 3     & & \\[4pt]
    \quad Observations                                  & 1444 & & \\[4pt]
    \hline
    \end{tabular}
    \end{table}


  \begin{table}[H]
    \caption{Zero-inflated lognormal GLMM for absolute perennial forb cover (\%).
    Model structure and interpretation follow Table~\ref{tab:zi_lognormal_annual}.}
    \label{tab:table4_fig2}
    \centering
    \begin{tabular}{lccc}
    \multicolumn{4}{c}{\textbf{Perennial forb cover (\%)}} \\[4pt]
    \hline
    \textit{Predictors} & \textit{Estimates} & \textit{CI} & \textit{p} \\
    \hline\\[-6pt]
        \textbf{Continuous Model} & & & \\[4pt]
    \quad (Intercept)              & 21.57 & 17.20 -- 27.05 & \textbf{$<$0.001} \\[4pt]
    \quad Disturbance              & 0.93  & 0.86 -- 1.02   & 0.124 \\[4pt]
    \quad NPK                      & 0.84  & 0.77 -- 0.92   & \textbf{$<$0.001} \\[4pt]
    \quad Disturbance $\times$ NPK & 1.15  & 1.02 -- 1.30   & \textbf{0.022} \\[4pt]
    \\[-6pt]
    \textbf{Zero-Inflated Model} & & & \\[4pt]
    \quad (Intercept)              & 0.03  & 0.01 -- 0.11   & \textbf{$<$0.001} \\[4pt]
    \quad Disturbance              & 1.10  & 0.68 -- 1.77   & 0.700 \\[4pt]
    \quad NPK                      & 2.31  & 1.46 -- 3.66   & \textbf{$<$0.001} \\[4pt]
    \quad Disturbance $\times$ NPK & 0.47  & 0.24 -- 0.92   & \textbf{0.027} \\[4pt]
    \\[-6pt]
    \textbf{Random Effects} & & & \\[4pt]
    \quad $\sigma^{2}$                                   & 28.44 & & \\[4pt]
    \quad $\tau_{00}$ \scriptsize{site\_name}            & 0.40  & & \\[4pt]
    \quad $\tau_{00}$ \scriptsize{site\_name:year\_trt}  & 0.06  & & \\[4pt]
    \quad ICC                                            & 0.02  & & \\[4pt]
    \quad N \scriptsize{site\_name}                      & 37    & & \\[4pt]
    \quad N \scriptsize{year\_trt}                       & 3     & & \\[4pt]
    \quad Observations                                  & 1444 & & \\[4pt]
    \hline
    \end{tabular}
    \end{table}


  \begin{table}[H]
    \caption{Zero-inflated lognormal GLMM for absolute annual legume cover (\%).
   Model structure and interpretation follow Table~\ref{tab:zi_lognormal_annual}.}
    \label{tab:table5_fig2}
    \centering
    \begin{tabular}{lccc}
    \multicolumn{4}{c}{\textbf{Annual legume cover (\%)}} \\[4pt]
    \hline
    \textit{Predictors} & \textit{Estimates} & \textit{CI} & \textit{p} \\
    \hline\\[-6pt]
        \textbf{Continuous Model} & & & \\[4pt]
    \quad (Intercept)              & 9.14     & 6.43 -- 12.98       & \textbf{$<$0.001} \\[4pt]
    \quad Disturbance              & 0.97     & 0.80 -- 1.19        & 0.785 \\[4pt]
    \quad NPK                      & 0.84     & 0.70 -- 1.01        & 0.069 \\[4pt]
    \quad Disturbance $\times$ NPK & 1.31     & 1.00 -- 1.72        & 0.050 \\[4pt]
    \\[-6pt]
    \textbf{Zero-Inflated Model} & & & \\[4pt]
    \quad (Intercept)              & 10239.36 & 442.00 -- 237207.08 & \textbf{$<$0.001} \\[4pt]
    \quad Disturbance              & 1.39     & 0.73 -- 2.66        & 0.315 \\[4pt]
    \quad NPK                      & 1.61     & 0.85 -- 3.06        & 0.146 \\[4pt]
    \quad Disturbance $\times$ NPK & 0.55     & 0.22 -- 1.37        & 0.201 \\[4pt]
    \\[-6pt]
    \textbf{Random Effects} & & & \\[4pt]
    \quad $\sigma^{2}$                                   & 21.96 & & \\[4pt]
    \quad $\tau_{00}$ \scriptsize{site\_name}            & 0.26  & & \\[4pt]
    \quad $\tau_{00}$ \scriptsize{site\_name:year\_trt}  & 0.02  & & \\[4pt]
    \quad ICC                                            & 0.01  & & \\[4pt]
    \quad N \scriptsize{site\_name}                      & 37    & & \\[4pt]
    \quad N \scriptsize{year\_trt}                       & 3     & & \\[4pt]
    \quad Observations                                  & 1444 & & \\[4pt]
    \hline
    \multicolumn{4}{p{11cm}}{\textit{Note:}   the odds-ratio scale (442.00--237207.08), this corresponds to an estimated zero-inflation probability between 0.998 and 0.999996.}
    \end{tabular}
    \end{table}

  
  \begin{table}[H]
    \caption{Zero-inflated lognormal GLMM for absolute perennial legume cover (\%).
   Model structure and interpretation follow Table~\ref{tab:zi_lognormal_annual}.}
    \label{tab:table6_fig2}
    \centering
    \begin{tabular}{lccc}
    \multicolumn{4}{c}{\textbf{Perennial legume cover (\%)}} \\[4pt]
    \hline
    \textit{Predictors} & \textit{Estimates} & \textit{CI} & \textit{p} \\
    \hline\\[-6pt]
        \textbf{Continuous Model} & & & \\[4pt]
    \quad (Intercept)              & 10.38 & 8.33 -- 12.92 & \textbf{$<$0.001} \\[4pt]
    \quad Disturbance              & 0.68  & 0.58 -- 0.80  & \textbf{$<$0.001} \\[4pt]
    \quad NPK                      & 0.96  & 0.81 -- 1.13  & 0.597 \\[4pt]
    \quad Disturbance $\times$ NPK & 1.14  & 0.91 -- 1.43  & 0.247 \\[4pt]
    \\[-6pt]
    \textbf{Zero-Inflated Model} & & & \\[4pt]
    \quad (Intercept)              & 5.85  & 1.79 -- 19.14 & \textbf{0.003} \\[4pt]
    \quad Disturbance              & 0.83  & 0.53 -- 1.31  & 0.434 \\[4pt]
    \quad NPK                      & 1.62  & 1.02 -- 2.58  & \textbf{0.041} \\[4pt]
    \quad Disturbance $\times$ NPK & 1.01  & 0.53 -- 1.94  & 0.973 \\[4pt]
    \\[-6pt]
    \textbf{Random Effects} & & & \\[4pt]
    \quad $\sigma^{2}$                                   & 18.75 & & \\[4pt]
    \quad $\tau_{00}$ \scriptsize{site\_name}            & 0.12  & & \\[4pt]
    \quad $\tau_{00}$ \scriptsize{site\_name:year\_trt}  & 0.04  & & \\[4pt]
    \quad ICC                                            & 0.01  & & \\[4pt]
    \quad N \scriptsize{site\_name}                      & 37    & & \\[4pt]
    \quad N \scriptsize{year\_trt}                       & 3     & & \\[4pt]
    \quad Observations                                  & 1444 & & \\[4pt]
    \hline
    \end{tabular}
    \end{table}




  \newcommand{\PrecTempInt}{Precipitation of Warmest Q $\times$ Mean Temp of Warmest Q }

\begin{table}[H]
\caption{Ordered beta mixed-effects regression for mean relative annual cover, combining Control and Disturbance observations ($n = 74$ site $\times$ treatment observations; 37 sites $\times$ 2 treatments), as a function of mean temperature and precipitation of the warmest quarter (both z-scored), Disturbance, and the temperature~$\times$~precipitation interaction, with a random intercept for site.
The table has three row groups:
\textit{Mu (Mean)}; logit-scale coefficients describing variation in annual prevalence;
\textit{Sigma (Precision)}; the beta precision parameter, held constant across climate and treatment;
and \textit{Cutpoints (Boundary)}; the estimated lower and upper cutpoints governing the occurrence of observations at the boundaries ($Y = 0$ and $Y = 1$).
The intercept in the Mu component gives the linear predictor at mean climate under Control.
This is the reduced model retained after the interaction model comparison reported in Table~\ref{tab:climate_interaction_lrt}.}
\label{tab:climate_cover_combined}
\centering
\begin{tabular}{p{7.5cm}ccc}
\toprule
\textit{Predictors} & \textit{Estimates} & \textit{CI} & \textit{p} \\
\midrule
\textbf{Mu (Mean)} & & & \\[3pt]
 (Intercept)                     & $-$1.84 & $-$2.40 -- $-$1.27 & \textbf{$<$0.001} \\[3pt]
 Precipitation of Warmest Q      & $-$1.29 & $-$1.84 -- $-$0.74 & \textbf{$<$0.001} \\[3pt]
 Mean Temp of Warmest Q          & 1.20    & 0.68 -- 1.72       & \textbf{$<$0.001} \\[3pt]
 Disturbance                     & 1.18    & 0.72 -- 1.63       & \textbf{$<$0.001} \\[3pt]
 \PrecTempInt                    & $-$0.22 & $-$0.77 -- 0.32    & 0.417 \\[3pt]
\midrule
\textbf{Sigma (Precision)} & & & \\[3pt]
 (Intercept)                     & 2.05    & 1.47 -- 2.63       & \textbf{$<$0.001} \\[3pt]
\midrule
\textbf{Cutpoints (Boundary)} & & & \\[3pt]
 Lower cutoff                    & $-$4.24 & $-$5.24 -- $-$3.25 & \textbf{$<$0.001} \\[3pt]
 Upper cutoff                    & 4.50    & 2.65 -- 6.36       & \textbf{$<$0.001} \\[3pt]
\midrule
\multicolumn{4}{l}{$N = 74$; AIC $= -34.1$} \\
\bottomrule
\end{tabular}
\end{table}

\begin{table}[H]
\caption{Ordered beta mixed-effects regression for mean annual species share, combining Control and Disturbance observations ($n = 74$ site $\times$ treatment observations; 37 sites $\times$ 2 treatments). Model structure and column definitions are identical to Table~\ref{tab:climate_cover_combined}, with annual species share (proportion of species that are annual) as the response instead of relative annual cover. This is the reduced model retained after the interaction model comparison reported in Table~\ref{tab:climate_interaction_lrt}.}
\label{tab:climate_species_combined}
\centering
\begin{tabular}{p{7.5cm}ccc}
\toprule
\textit{Predictors} & \textit{Estimates} & \textit{CI} & \textit{p} \\
\midrule
\textbf{Mu (Mean)} & & & \\[3pt]
 (Intercept)                & $-$1.50 & $-$1.93 -- $-$1.07 & \textbf{$<$0.001} \\[3pt]
 Precipitation of Warmest Q & $-$1.06 & $-$1.53 -- $-$0.59 & \textbf{$<$0.001} \\[3pt]
 Mean Temp of Warmest Q     & 0.98    & 0.54 -- 1.41       & \textbf{$<$0.001} \\[3pt]
 Disturbance                & 0.82    & 0.54 -- 1.10       & \textbf{$<$0.001} \\[3pt]
 \PrecTempInt               & $-$0.18 & $-$0.66 -- 0.29    & 0.451 \\[3pt]
\midrule
\textbf{Sigma (Precision)} & & & \\[3pt]
 (Intercept)                & 2.97    & 2.49 -- 3.46       & \textbf{$<$0.001} \\[3pt]
\midrule
\textbf{Cutpoints (Boundary)} & & & \\[3pt]
 Lower cutoff               & $-$3.82 & $-$4.67 -- $-$2.97 & \textbf{$<$0.001} \\[3pt]
 Upper cutoff               & 4.02    & 2.36 -- 5.67       & \textbf{$<$0.001} \\[3pt]
\midrule
\multicolumn{4}{l}{$N = 74$; AIC $= -12.6$} \\
\bottomrule
\end{tabular}
\end{table}

\begin{table}[H]
\caption{Model comparison between the full climate $\times$ disturbance interaction model and the reduced (additive-disturbance) ordered beta model, for relative annual cover and annual species share (Control and Disturbance combined, $n = 74$).
\textbf{Full}: response $\sim$ Precipitation $\times$ Temperature $\times$ Disturbance $+ (1\mid\text{site})$.
\textbf{Reduced}: response $\sim$ Precipitation $\times$ Temperature $+$ Disturbance $+ (1\mid\text{site})$, i.e.\ no Disturbance $\times$ climate interaction terms.
Likelihood ratio tests (LRT) indicate no improvement in fit from the interaction terms for either response, supporting the reduced model used to generate Figure~\ref{fig:climate_fig4}.}
    \label{tab:climate_interaction_lrt}
    \centering
    \begin{tabular}{lccccccc}
    \toprule
    \textit{Model} & \textit{df} & \textit{AIC} & \textit{BIC} & \textit{logLik} & \textit{deviance} & \textit{$\chi^2$ (df)} & \textit{p} \\
    \midrule
    \multicolumn{8}{l}{\textbf{Relative annual cover}} \\[3pt]
    \quad Reduced & 9  & $-$34.08 & $-$13.35 & 26.04 & $-$52.08 & --       & --    \\[3pt]
    \quad Full    & 12 & $-$30.06 & $-$2.41  & 27.03 & $-$54.06 & 1.97 (3) & 0.578 \\[6pt]
    \multicolumn{8}{l}{\textbf{Annual species share}} \\[3pt]
    \quad Reduced & 9  & $-$12.61 & 8.13  & 15.31 & $-$30.61 & --       & --    \\[3pt]
    \quad Full    & 12 & $-$7.14  & 20.51 & 15.57 & $-$31.14 & 0.53 (3) & 0.913 \\[3pt]
    \bottomrule
    \end{tabular}
    \end{table}

\begin{table}[H]
\footnotesize
    \caption{Ordered beta mixed-effects regression of relative annual cover on mean temperature and precipitation of the warmest quarter, Disturbance, and their interactions (full factorial model, Control and Disturbance fitted jointly with a random intercept for site).
    Estimates are exponentiated and describe multiplicative effects on the odds scale; values greater than 1 indicate higher and values below 1 indicate lower annual cover. Climate variables are $z$-scored.
    Disturbance~$\times$~climate terms test whether the effect of Disturbance depends on climate; none were significant.
    $\sigma^{2}$ is the residual variance, $\tau_{00}$ the site-level random-intercept variance, and ICC the intraclass correlation. $N_{\text{site}} = 37$.}
    \label{tab:climate_interaction_wald_cover}
    \centering
    \begin{tabular}{lccc}
    \multicolumn{4}{c}{\textbf{Annual cover}} \\[4pt]
    \hline
    \textit{Predictors} & \textit{Estimates} & \textit{CI} & \textit{p} \\
    \hline\\[-6pt]
    (Intercept) & 0.15 & 0.08 -- 0.27 & \textbf{$<$0.001} \\[4pt]
    Precipitation (warmest quarter) & 0.28 & 0.15 -- 0.53 & \textbf{$<$0.001} \\[4pt]
    Temperature (warmest quarter) & 3.83 & 2.04 -- 7.16 & \textbf{$<$0.001} \\[4pt]
    Disturbance & 3.18 & 1.99 -- 5.09 & \textbf{$<$0.001} \\[4pt]
    Precipitation $\times$ Temperature & 0.68 & 0.35 -- 1.31 & 0.253 \\[4pt]
    Precipitation $\times$ Disturbance & 0.92 & 0.55 -- 1.54 & 0.750 \\[4pt]
    Temperature $\times$ Disturbance & 0.81 & 0.50 -- 1.31 & 0.390 \\[4pt]
    Precipitation $\times$ Temperature $\times$ Disturbance & 1.33 & 0.77 -- 2.29 & 0.303 \\[4pt]
    \\[-6pt]
    \textbf{Random Effects} & & & \\[4pt]
    \quad $\sigma^{2}$ & 0.26 & & \\[4pt]
    \quad $\tau_{00}$ \scriptsize{site\_name} & 1.45 & & \\[4pt]
    \quad ICC & 0.85 & & \\[4pt]
    \quad $N_{\text{site}}$ & 37 & & \\[4pt]
    \hline\\[-6pt]
    Observations & 74 & & \\[4pt]
    Marginal $R^{2}$ / Conditional $R^{2}$ & 0.676 / 0.951 & & \\[4pt]
    \hline
    \end{tabular}
\end{table}

\begin{table}[H]
\footnotesize
    \caption{Ordered beta mixed-effects regression of the proportion of annual species on mean temperature and precipitation of the warmest quarter, Disturbance, and their interactions (full factorial model, Control and Disturbance fitted jointly with a random intercept for site).
    Estimates are exponentiated and describe multiplicative effects on the odds scale; values greater than 1 indicate a higher and values below 1 a lower proportion of annual species. Climate variables are $z$-scored.
    Disturbance~$\times$~climate terms test whether the effect of Disturbance depends on climate; none were significant.
    $\sigma^{2}$ is the residual variance, $\tau_{00}$ the site-level random-intercept variance, and ICC the intraclass correlation. $N_{\text{site}} = 37$.}
    \label{tab:climate_interaction_wald_species}
    \centering
    \begin{tabular}{lccc}
    \multicolumn{4}{c}{\textbf{Annual species}} \\[4pt]
    \hline
    \textit{Predictors} & \textit{Estimates} & \textit{CI} & \textit{p} \\
    \hline\\[-6pt]
    (Intercept) & 0.21 & 0.14 -- 0.34 & \textbf{$<$0.001} \\[4pt]
    Precipitation (warmest quarter) & 0.32 & 0.19 -- 0.54 & \textbf{$<$0.001} \\[4pt]
    Temperature (warmest quarter) & 2.74 & 1.69 -- 4.47 & \textbf{$<$0.001} \\[4pt]
    Disturbance & 2.38 & 1.74 -- 3.27 & \textbf{$<$0.001} \\[4pt]
    Precipitation $\times$ Temperature & 0.90 & 0.53 -- 1.52 & 0.690 \\[4pt]
    Precipitation $\times$ Disturbance & 1.13 & 0.78 -- 1.63 & 0.522 \\[4pt]
    Temperature $\times$ Disturbance & 0.96 & 0.69 -- 1.35 & 0.823 \\[4pt]
    Precipitation $\times$ Temperature $\times$ Disturbance & 0.88 & 0.60 -- 1.30 & 0.524 \\[4pt]
    \\[-6pt]
    \textbf{Random Effects} & & & \\[4pt]
    \quad $\sigma^{2}$ & 0.12 & & \\[4pt]
    \quad $\tau_{00}$ \scriptsize{site\_name} & 1.05 & & \\[4pt]
    \quad ICC & 0.89 & & \\[4pt]
    \quad $N_{\text{site}}$ & 37 & & \\[4pt]
    \hline\\[-6pt]
    Observations & 74 & & \\[4pt]
    Marginal $R^{2}$ / Conditional $R^{2}$ & 0.654 / 0.963 & & \\[4pt]
    \hline
    \end{tabular}
\end{table}

\newpage
  \setlength\LTpre{0pt}
  \setlength\LTpost{0pt}

   \begin{longtable}{@{}llrrrr@{}}
\caption{Model selection for climate predictors using ordered beta regression. Each row is one combination of a temperature variable and a precipitation (or aridity) variable, fitted to Control treatment data only for both mean relative annual cover and mean annual species share across sites ($n = 37$). All predictors were $z$-scored before fitting. Models include a temperature $\times$ precipitation interaction; precision was held constant. $\Delta\mathrm{AIC}$ is the difference from the lowest AIC within each response column; values $<$ 2 indicate similar fit. Mean temperature of the warmest quarter (TEMP\_WARM\_Q) and precipitation of the warmest quarter (MAP\_WARM\_Q) achieved the lowest AIC for both responses and were retained as the climatic predictors in the subsequent Control--Disturbance analyses. Rows are ordered by $\Delta\mathrm{AIC}_\text{cov}$.
Abbreviations: MAT $=$ mean annual temperature; AI $=$ aridity index; PET $=$ potential evapotranspiration; Q $=$ climatic quarter (3-month period).}
    \label{tab:climate_long_table}\\
    \toprule
    \textit{Temperature} & \textit{Precipitation} & \textit{AIC$_\text{cov}$} & \textit{AIC$_\text{sp}$} & \textit{$\Delta$AIC$_\text{cov}$} & \textit{$\Delta$AIC$_\text{sp}$} \\
    \midrule
    \endfirsthead
    \multicolumn{6}{l}{\small\textit{Table~\ref{tab:climate_long_table} continued}}\\[2pt]
    \toprule
    \textit{Temperature} & \textit{Precipitation} &
    \textit{AIC$_\text{cov}$} & \textit{AIC$_\text{sp}$} &
    \textit{$\Delta$AIC$_\text{cov}$} & \textit{$\Delta$AIC$_\text{sp}$} \\
    \midrule
    \endhead
    \midrule
    \multicolumn{6}{r}{\footnotesize Continued on next page}\\
    \endfoot
    \bottomrule
    \endlastfoot

    \textbf{TEMP\_WARM\_Q} & \textbf{MAP\_WARM\_Q} & \textbf{$-$12.4} & \textbf{13.8} & \textbf{0.0} & \textbf{0.0} \\
    TEMP\_WET\_Q & MAP\_WARM\_Q & $-$10.3 & 14.8 & 2.1 & 1.1 \\
    MAT & MAP\_WARM\_Q & $-$7.5 & 14.4 & 4.9 & 0.6 \\
    TEMP\_DRY\_Q & MAP\_WARM\_Q & $-$4.4 & 17.4 & 8.1 & 3.7 \\
    TEMP\_COLD\_Q & MAP\_WARM\_Q & $-$2.0 & 17.5 & 10.5 & 3.8 \\
    TEMP\_DRY\_Q & MAP\_DRY\_Q & $-$1.5 & 19.8 & 10.9 & 6.0 \\
    TEMP\_WARM\_Q & MAP\_DRY\_Q & $-$1.1 & 23.3 & 11.3 & 9.5 \\
    MIN\_TEMP & MAP\_WARM\_Q & $-$0.2 & 19.0 & 12.2 & 5.3 \\
    TEMP\_DRY\_Q & AI & 0.8 & 23.8 & 13.3 & 10.1 \\
    MAT & AI & 1.5 & 24.9 & 14.0 & 11.1 \\
    TEMP\_WARM\_Q & MAP & 1.7 & 26.5 & 14.1 & 12.8 \\
    TEMP\_WARM\_Q & AI & 1.8 & 25.6 & 14.3 & 11.8 \\
    TEMP\_COLD\_Q & AI & 1.9 & 25.0 & 14.3 & 11.2 \\
    MAT\_RANGE & AI & 2.0 & 25.0 & 14.4 & 11.2 \\
    MAT\_RANGE & MAP\_WARM\_Q & 2.1 & 23.8 & 14.5 & 10.1 \\
    MIN\_TEMP & AI & 2.1 & 25.1 & 14.5 & 11.3 \\
    TEMP\_WET\_Q & MAP\_DRY\_Q & 2.5 & 24.9 & 14.9 & 11.2 \\
    MAT & MAP\_DRY\_Q & 2.9 & 24.3 & 15.3 & 10.6 \\
    TEMP\_WET\_Q & PET & 3.0 & 25.3 & 15.5 & 11.5 \\
    ANN\_TEMP\_RANGE & MAP\_WARM\_Q & 3.1 & 24.4 & 15.5 & 10.6 \\
    TEMP\_VAR & MAP\_WARM\_Q & 3.2 & 24.0 & 15.7 & 10.2 \\
    TEMP\_DRY\_Q & MAP & 3.3 & 23.6 & 15.7 & 9.9 \\
    ANN\_TEMP\_RANGE & AI & 4.1 & 26.1 & 16.6 & 12.3 \\
    MIN\_TEMP & MAP\_DRY\_Q & 4.3 & 24.9 & 16.8 & 11.1 \\
    TEMP\_WET\_Q & MAP\_VAR & 4.4 & 24.0 & 16.8 & 10.2 \\
    TEMP\_COLD\_Q & MAP\_DRY\_Q & 4.4 & 25.0 & 16.8 & 11.2 \\
    MAT & MAP & 4.5 & 26.6 & 17.0 & 12.8 \\
    TEMP\_WET\_Q & AI & 4.6 & 27.2 & 17.0 & 13.4 \\
    TEMP\_VAR & AI & 4.6 & 27.1 & 17.0 & 13.4 \\
    TEMP\_DRY\_Q & MAP\_COLD\_Q & 4.9 & 25.3 & 17.3 & 11.6 \\
    ANN\_TEMP\_RANGE & MAP\_DRY\_Q & 5.4 & 27.2 & 17.8 & 13.4 \\
    MAT\_RANGE & MAP\_DRY\_Q & 5.6 & 28.3 & 18.0 & 14.5 \\
    TEMP\_VAR & MAP\_DRY\_Q & 5.7 & 27.3 & 18.1 & 13.6 \\
    TEMP\_DRY\_Q & PET & 6.1 & 28.4 & 18.6 & 14.6 \\
    TEMP\_DRY\_Q & MAP\_VAR & 6.3 & 25.8 & 18.7 & 12.1 \\
    TEMP\_COLD\_Q & MAP & 7.5 & 28.3 & 19.9 & 14.6 \\
    TEMP\_VAR & MAP\_VAR & 7.7 & 30.2 & 20.1 & 16.4 \\
    TEMP\_WARM\_Q & MAP\_VAR & 8.0 & 29.8 & 20.4 & 16.0 \\
    MIN\_TEMP & MAP & 8.1 & 29.2 & 20.5 & 15.4 \\
    TEMP\_DRY\_Q & MAP\_WET\_Q & 8.1 & 28.2 & 20.5 & 14.5 \\
    MAT\_RANGE & PET & 8.3 & 29.6 & 20.8 & 15.8 \\
    ANN\_TEMP\_RANGE & MAP\_VAR & 8.6 & 30.4 & 21.0 & 16.6 \\
    MAT\_RANGE & MAP\_VAR & 8.9 & 30.1 & 21.4 & 16.4 \\
    TEMP\_WARM\_Q & MAP\_WET\_Q & 9.2 & 33.0 & 21.6 & 19.3 \\
    MIN\_TEMP & MAP\_VAR & 9.2 & 30.5 & 21.7 & 16.8 \\
    TEMP\_COLD\_Q & MAP\_VAR & 9.3 & 30.6 & 21.8 & 16.8 \\
    MAT & MAP\_VAR & 9.5 & 30.5 & 21.9 & 16.7 \\
    MAT & PET & 10.7 & 33.6 & 23.2 & 19.9 \\
    TEMP\_COLD\_Q & PET & 11.2 & 34.4 & 23.6 & 20.6 \\
    MAT\_RANGE & MAP & 11.4 & 35.0 & 23.8 & 21.3 \\
    TEMP\_VAR & PET & 11.5 & 34.6 & 23.9 & 20.9 \\
    TEMP\_WET\_Q & MAP & 11.7 & 35.0 & 24.1 & 21.3 \\
    ANN\_TEMP\_RANGE & MAP & 11.7 & 34.5 & 24.2 & 20.8 \\
    TEMP\_VAR & MAP & 11.8 & 34.2 & 24.2 & 20.4 \\
    MIN\_TEMP & PET & 11.8 & 34.6 & 24.2 & 20.9 \\
    TEMP\_COLD\_Q & MAP\_COLD\_Q & 11.9 & 36.2 & 24.4 & 22.4 \\
    MAT & MAP\_WET\_Q & 12.1 & 33.7 & 24.5 & 20.0 \\
    ANN\_TEMP\_RANGE & PET & 12.3 & 34.5 & 24.8 & 20.7 \\
    TEMP\_WARM\_Q & PET & 12.4 & 34.2 & 24.8 & 20.5 \\
    MIN\_TEMP & MAP\_COLD\_Q & 14.1 & 37.5 & 26.5 & 23.7 \\
    TEMP\_COLD\_Q & MAP\_WET\_Q & 15.1 & 36.3 & 27.5 & 22.5 \\
    MAT & MAP\_COLD\_Q & 15.1 & 39.0 & 27.6 & 25.2 \\
    TEMP\_WARM\_Q & MAP\_COLD\_Q & 15.6 & 39.3 & 28.0 & 25.5 \\
    MIN\_TEMP & MAP\_WET\_Q & 15.8 & 37.5 & 28.2 & 23.7 \\
    MAT\_RANGE & MAP\_WET\_Q & 17.7 & 41.3 & 30.2 & 27.5 \\
    ANN\_TEMP\_RANGE & MAP\_WET\_Q & 17.8 & 41.1 & 30.3 & 27.3 \\
    TEMP\_VAR & MAP\_WET\_Q & 17.9 & 41.2 & 30.3 & 27.4 \\
    TEMP\_VAR & MAP\_COLD\_Q & 18.0 & 41.6 & 30.4 & 27.8 \\
    TEMP\_WET\_Q & MAP\_WET\_Q & 18.1 & 41.2 & 30.6 & 27.5 \\
    ANN\_TEMP\_RANGE & MAP\_COLD\_Q & 19.2 & 42.7 & 31.6 & 29.0 \\
    MAT\_RANGE & MAP\_COLD\_Q & 19.7 & 43.1 & 32.1 & 29.3 \\
    TEMP\_WET\_Q & MAP\_COLD\_Q & 19.8 & 42.4 & 32.2 & 28.6 \\

  \end{longtable}

\end{document}